\documentclass[fp,seceq,twocolumn]{jpsj3}
\usepackage{txfonts,bm,amsmath,amssymb,url,float,cases,braket,xcolor,graphicx}

\renewcommand{\H}{{\mathcal H}}

\newcommand{\bS}{{\bm{S}}}

\newcommand{\bq}{{\bm{q}}}
\newcommand{\bp}{{\bm{p}}}
\newcommand{\bP}{{\bm{P}}}

\newcommand{\TN}{T_{\rm N}}

\newcommand{\Co}{{Co$_4$Nb$_2$O$_9$}}
\newcommand{\Sr}{{Sr$_2$CoGe$_2$O$_7$}}
\newcommand{\Ba}{{Ba$_2$CoGe$_2$O$_7$}}
\newcommand{\Tl}{{TlCuCl$_3$}}

\newcommand{\Tb}{{TbMnO$_3$}}

\renewcommand{\H}{{\mathcal H}}

\newcommand{\bH}{{\bm{H}}}
\newcommand{\bE}{{\bm{E}}}
\newcommand{\be}{{\bm{e}}}

\newcommand{\tbe}{\tilde{\bm{e}}}
\newcommand{\tx}{\tilde{x}}
\newcommand{\ty}{\tilde{y}}
\newcommand{\tz}{\tilde{z}}
\newcommand{\tS}{\tilde{S}}
\newcommand{\tp}{\tilde{p}}
\newcommand{\tK}{\tilde{K}}
\newcommand{\tO}{\tilde{O}}

\newcommand{\rG}{{\rm G}}
\newcommand{\rL}{{\rm L}}
\newcommand{\rT}{{\rm T}}

\newcommand{\rA}{{\rm A}}
\newcommand{\rB}{{\rm B}}

\newcommand{\Pc}{{$P\bar{3}c1$}}

\begin{document}

%%%%%%%%%%%%%%%%%%%%%%%%%%%%%%%%%%%%%%%%%%%%%%%%%%%%%%%%%%%%%%%%%%%%%%%%%%%%%%%%%%%%%%
\title{
Symmetry Analysis of Magnetoelectric Effects \\
in Honeycomb Antiferromagnet \Co
}

\author{Masashige Matsumoto$^1$\thanks{E-mail address: matsumoto.masashige@shizuoka.ac.jp} and Mikito Koga$^2$}

\inst{
$^1$Department of Physics, Faculty of Science, Shizuoka University, Shizuoka 422-8529, Japan \\
$^2$Department of Physics, Faculty of Education, Shizuoka University, Shizuoka 422-8529, Japan
}

\recdate{February 28, 2019}

%%%%%%%%%%%%%%%%%%%%%%%%%%%%%%%%%%%%%%%%%%%%%%%%%%%%%%%%%%%%%%%%%%%%%%%%%%%%%%%%%%%%%%

%%%%%%%%%%%%%%%%%%%%%%%%%%%%%%%%%%%%%%%%%%%%%%%%%%%%%%%%%%%%%%%%%%%%%%%%%%%%%%%%%%%%%%
\abst{
Magnetoelectric effects in honeycomb antiferromagnet \Co~are investigated
on the basis of symmetry analyses of Co$^{2+}$ ions in trigonal \Pc~space group.
For each Co$^{2+}$ ion, the possible spin dependence is classified by $C_3$ point-group symmetry.
This accounts for the observed main effect that an electric polarization rotates in the opposite direction at the twice speed
relative to the rotation of the external magnetic field applied in the $ab$-plane.
Inversion centers and twofold axes in the unit cell restrict the active spin-dependence of the electric polarization,
which well  explains the observed experimental results.
Expected optical properties of quadrupolar excitation and various types of dichroism are also discussed.
}
%%%%%%%%%%%%%%%%%%%%%%%%%%%%%%%%%%%%%%%%%%%%%%%%%%%%%%%%%%%%%%%%%%%%%%%%%%%%%%%%%%%%%%

\maketitle

%%%%%%%%%%%%%%%%%%%%%%%%%%%%%%%%%%%%%%%%%%%%%%%%%%%%%%%%%%%%%%%%%%%%%%%%%%%%%%%%%%%%%%%%%%%%%%%%%%%
\section{Introduction}
%%%%%%%%%%%%%%%%%%%%%%%%%%%%%%%%%%%%%%%%%%%%%%%%%%%%%%%%%%%%%%%%%%%%%%%%%%%%%%%%%%%%%%%%%%%%%%%%%%%

We propose that the honeycomb antiferromagnet \Co
\cite{Bertaut-1961}
is a typical example for the investigation of the magnetoelectric effects from a picture of quantum spin systems.
The magnetoelectric effects in A$_4$B$_2$O$_9$ (A: Co or Mn, B: Nb or Ta) system were found by Fischer and coworkers a long time ago.
\cite{Fischer-1972}
Since the reports on a spin-flop driven magneto-dielectric effect
\cite{Kolodiazhnyi-2011}
and a large magnetoelectric coupling in \Co,
\cite{Fang-2014}
this system attracted much attention.
\cite{Cao-2015,Khanh-2016,Solovyev-2016,Lu-2016,Khanh-2017,Deng-2018,Yanagi-2018-1,Yanagi-2018-2,Xie-2018}
Among them, Khanh and coworkers investigated the linear magnetoelectric effects in detail
and found that the electric polarization rotates in the opposite direction at the twice speed
relative to the rotation of the external magnetic field applied in the basal $ab$-plane.
\cite{Khanh-2016,Khanh-2017}
This point was microscopically studied by Yanagi and coworkers
from a band picture under a strong antiferromagnetic (AF) molecular field on the basis of the linear response theory.
\cite{Yanagi-2018-1,Yanagi-2018-2}
They revealed that the local spin-orbit interaction at the Co site and the honeycomb lattice structure
play an important role in the linear magnetoelectric effect.

The research in multiferroic materials has achieved major progress since the discovery of the large magnetoelectric effect in \Tb.
\cite{Kimura-2003}
In the sinusoidal magnetic ordered state, such as in \Tb,
it was revealed that the inverse Dzyaloshinskii-Moriya effect (or spin-current mechanism) is the origin of the ferroelectricity.
Here, the electric dipole operator is described by the product of spin operators at different sites.
\cite{Katsura-2005,Sergienko-2006}
Another major origin of the spin-dependent electric dipole is the metal-ligand hybridization mechanism,
where the electric dipole is described by the product of spin operators at the same site.
It is expected for a magnetic ion which occupies a site lacking the inversion symmetry.
It successfully explains the ferroelectric polarization in delafossite compounds Cu(Fe,Al)O$_2$.
\cite{Arima-2007}
The theory is also applicable to the magnetic-field-controlled electric polarization
\cite{Murakawa-2010,Murakawa-2012}
and optical properties including directional dichroism in \Ba.
\cite{Penc-2012,Kezsmarki-2011,Miyahara-2011,Kezsmarki-2014}

As mentioned above, there are two types of spin-dependent electric dipole.
One is described by product of spin operators at different sites (type-I),
whereas the other is by the same site (type-II).
Besides the microscopic origins of the spin-dependent electric dipole,
the possible spin dependences were classified by symmetries for the type-I
\cite{Kaplan-2011,Matsumoto-2017}
and the type-II.
\cite{Mims-1976,Matsumoto-2017}
The advantage of the theory is that the spin dependence in the electric dipole operator can be obtained
only by considering symmetries of the magnetic ions without going into its microscopic origin.
In particular, it demonstrates power for complicated crystal structures and enables us to analyze observed results.
For the type-I theory, it was applied to a spin dimer system
and revealed that the Bose-Einstein condensation of magnons induces a ferroelectricity in \Tl.
\cite{Kimura-2016,Kimura-2017}
It also explains forbidden transitions from singlet to triplet states for electron spin resonance.
\cite{Kimura-2018}

On the linear magnetoelectric effect in \Co,
there are theoretical works from the first principal calculation
\cite{Solovyev-2016}
and from the band picture
\cite{Yanagi-2018-1,Yanagi-2018-2}
in the presence of the spin-orbit interaction.
So far, no theoretical work has been reported on the basis of a quantum spin system.
In this paper, we study the magnetoelectric effects from the localized spin picture and apply the type-II theory to \Co,
where a quadrupole operator (product of spin operators at the same site) induces the electric dipole.
It is owing to the fact that the magnetic Co$^{2+}$ ion in \Co~occupies a site lacking the inversion symmetry
and both quadrupole and electric dipole can be classified in the same irreducible representation.
We derive the spin-dependent electric dipole operator at each Co site by means of the symmetry analysis of \Co~and show
that the averaged value of the electric dipole accounts for the observed electric polarization.
It is important to present the spin-based theory for the magnetoelectric effect in \Co.
Indeed, the symmetry analysis is easy to handle without going into details
and we can capture the essence of the magnetoelectric effects in terms of the quadrupole.
When we obtain the spin-dependent electric dipole operator, we can use it to elucidate related magnetoelectric effects such as optical properties for experiments.
As the expected optical effect in \Co, we discuss spin-quadrupolar excitations
\cite{Akaki-2017}
and various types of dichroism.

This paper is organized as follows.
In Sec. 2, we derive the spin-dependent electric dipole operator on the basis of the \Pc~space group
and show that the rotation of the electric polarization is the universal property
in the presence of the threefold rotational symmetry.
In Sec. 3, the spin-based theory is applied to optical effects
and discuss possible spin-quadrupolar excitation and observable dichroism in \Co.
The last section gives summary and discussions.

%%%%%%%%%%%%%%%%%%%%%%%%%%%%%%%%%%%%%%%%%%%%%%%%%%%%%%%%%%%%%%%%%%%%%%%%%%%%%%%%%%%%%%%%%%%%%%%%%%%
\section{Spin-Dependent Electric Dipole}
%%%%%%%%%%%%%%%%%%%%%%%%%%%%%%%%%%%%%%%%%%%%%%%%%%%%%%%%%%%%%%%%%%%%%%%%%%%%%%%%%%%%%%%%%%%%%%%%%%%

%%%%%%%%%%%%%%%%%%%%%%%%%%%%%%%%%%%%%%%%%%%%%%%%%%%%%%%%%%%%%%%%%%%%%%%%%%%%%%%%%%%%%%%%%%%%%%%%%%%
\subsection{Crystal structure and AF order}
%%%%%%%%%%%%%%%%%%%%%%%%%%%%%%%%%%%%%%%%%%%%%%%%%%%%%%%%%%%%%%%%%%%%%%%%%%%%%%%%%%%%%%%%%%%%%%%%%%%

%%%%%%%%%%%%%%%%%%%%%%%%%%%%%%%%%%%%%%%%%%%%%%%%%%%%%%%%%%%%%%%%%%%%%%%%%%%%%%%%%%%%%%
\begin{figure}[t]
\begin{center}
\includegraphics[width=6.5cm]{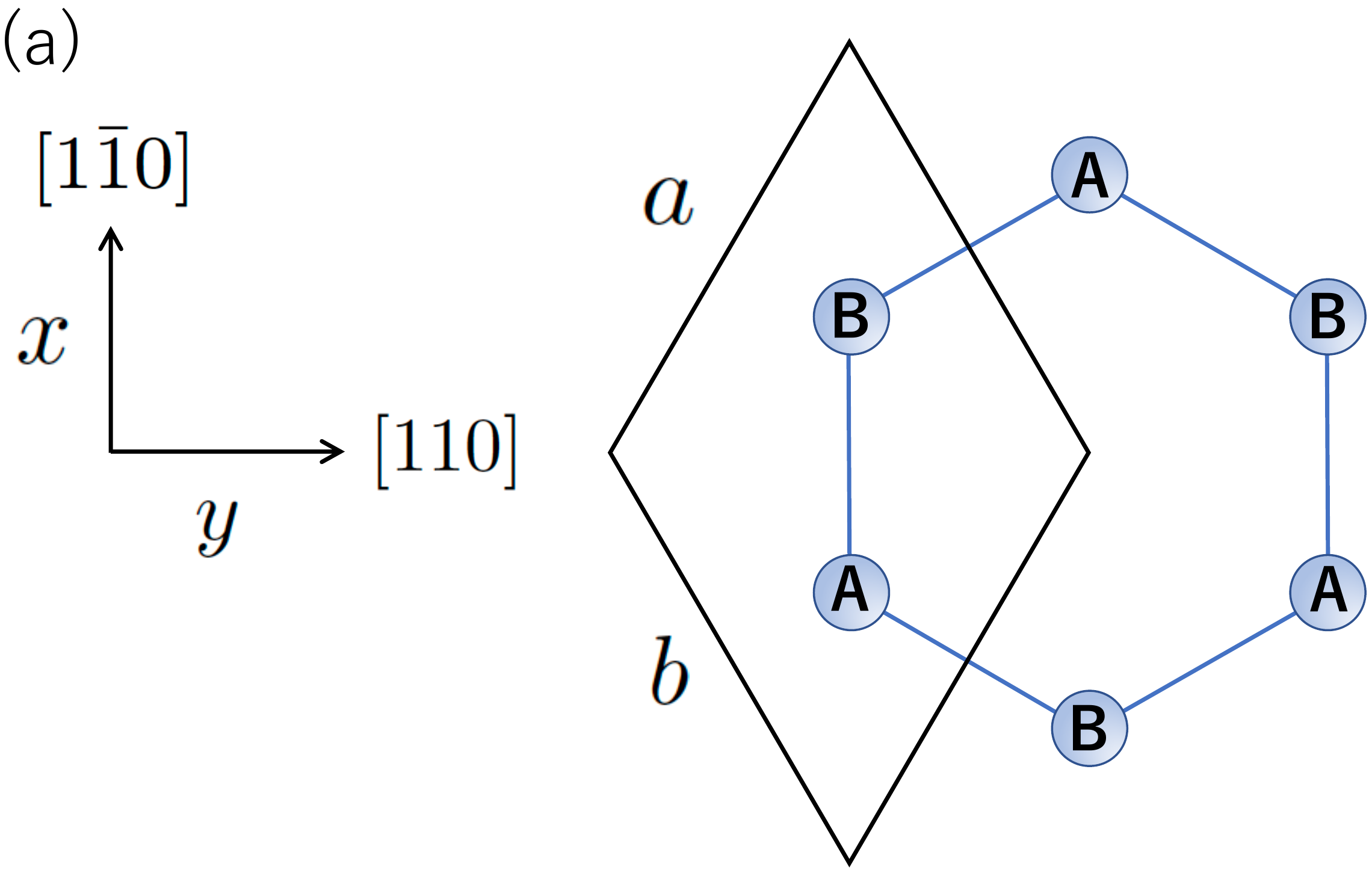}
\includegraphics[width=4cm]{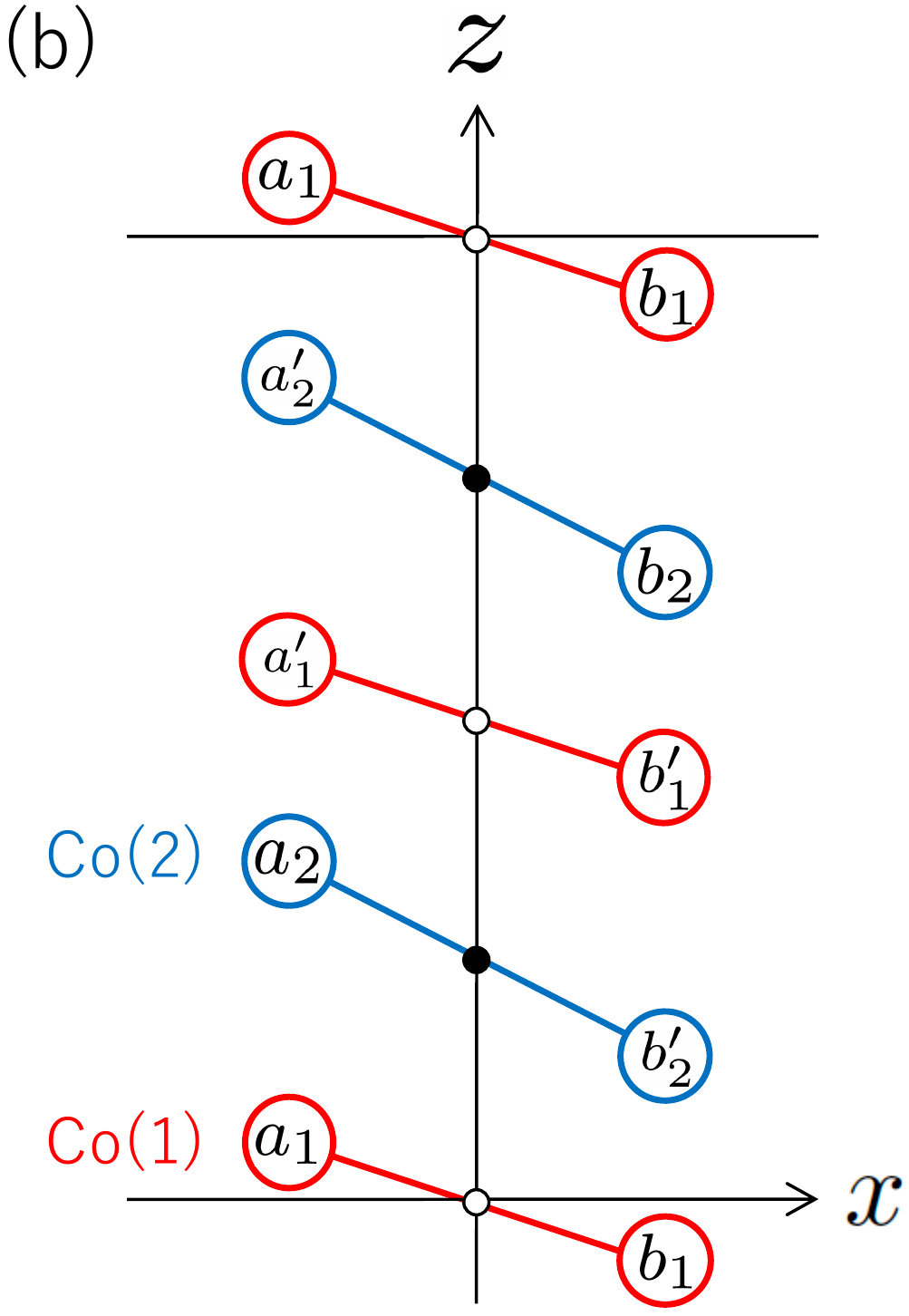}
\end{center}
\caption{
(Color online)
Schematic of crystal structure of \Co.
\cite{Bertaut-1961,Castellanos-2006,Deng-2018}
(a) Honeycomb lattice structure formed by Co$^{2+}$ ions (blue circles).
We take $x$, $y$, and $z$ axes along the $[1\bar{1}0]$, $[110]$, and $[001]$ directions, respectively.
The rhombus represents the unit cell.
A and B are two Co sites in a regular hexagon.
In the AF ordered phase, the ordered moment aligns along the $[1\bar{1}0]$ direction.
The moment is staggered on the A and B sites.
The local symmetry at the Co sites is represented by $C_3$ point group.
(b) Non-equivalent Co(1) and Co(2) sites in a unit cell.
There are four equivalent $(a_1, b_1, a'_1, b'_1)$ and $(a_2, b_2, a'_2, b'_2)$ sites
for the Co(1) and Co(2), respectively.
Here, $a$ and $b$ correspond to the A(up) and B(down) sites in the AF phase, respectively.
The Co$^{2+}$ ions connected by solid lines form honeycomb lattices at various positions of $z$.
The honeycomb lattices are not flat but are buckled in the $z$ direction.
We have four honeycomb lattices in the unit cell along the $z$ direction.
There are inversion centers and twofold axes in a unit cell for the trigonal \Pc~space group.
In the $z$-axis, each small white circle represents the inversion center,
while each small black circle represents the twofold axis along the $y$ direction.
}
\label{fig:crystal}
\end{figure}
%%%%%%%%%%%%%%%%%%%%%%%%%%%%%%%%%%%%%%%%%%%%%%%%%%%%%%%%%%%%%%%%%%%%%%%%%%%%%%%%%%%%%%

A schematic of crystal structure of \Co~is shown in Fig. \ref{fig:crystal}.
\cite{Bertaut-1961,Castellanos-2006,Deng-2018}
Co$^{2+}$ ions form a honeycomb lattice structure along the $ab$-plane through $O^{2-}$ ions.
The space group is \Pc~and the local symmetry at the Co site is represented by the $C_3$ point group with no inversion symmetry at the Co site.
\cite{Bertaut-1961,Deng-2018,Castellanos-2006}
Non-equivalent Co sites are denoted by Co(1) and Co(2).
In the trigonal \Pc~space group, there are inversion centers and twofold axes in a unit cell.
\cite{IN-table}
For the Co($i$) ($i=1,2$) site, $a_i\leftrightarrow b_i$ and $a'_i\leftrightarrow b'_i$ sites are related by the inversion center,
while $a_i\leftrightarrow b'_i$ and $b_i\leftrightarrow a'_i$ sites are related by the twofold axis
[see Fig. \ref{fig:crystal}(b)].
Along the $c$-axis, there are four honeycomb lattices in a unit cell.
We have eight Co$^{2+}$ ions in total in the unit cell.
The electric polarization is obtained by adding the all contributions from the eight ions.
The local $C_3$ point-group symmetry, inversion centers, and twofold axes are the keys to understanding the magnetoelectric effects in \Co.

In \Co, the magnetic Co$^{2+}$ ion carries $S=3/2$.
It was reported that  \Co~undergoes an AF phase transition below the N\'{e}el temperature $\TN=27.4$ K.
\cite{Bertaut-1961,Kolodiazhnyi-2011}
Neutron diffraction measurements revealed that the AF moment almost lies parallel to the $[1\bar{1}0]$ direction in the basal $ab$-plane,
however, there is a minor discrepancy between the single crystal and powder samples.
The former reported that the AF moment slightly canted towards the $c$-axis from the $ab$-plane,
\cite{Khanh-2016}
whereas the latter reported that the moment slightly canted in the $ab$-plane and it showed a noncollinear structure.
\cite{Deng-2018}
Since the main common result is that the AF moment aligns in the basal $ab$-plane,
we assume this structure to capture the essence of the magnetoelectric effects in \Co.

%%%%%%%%%%%%%%%%%%%%%%%%%%%%%%%%%%%%%%%%%%%%%%%%%%%%%%%%%%%%%%%%%%%%%%%%%%%%%%%%%%%%%%%%%%%%%%%%%%%
\subsection{Spin-dependent electric dipole in $C_3$ symmetry}
\label{sec:C3}
%%%%%%%%%%%%%%%%%%%%%%%%%%%%%%%%%%%%%%%%%%%%%%%%%%%%%%%%%%%%%%%%%%%%%%%%%%%%%%%%%%%%%%%%%%%%%%%%%%%

Let us consider a magnetic ion inducing an electric dipole.
Since the electric dipole is even with respect to the time-reversal transformation,
it is described by product of spin operators.
The spin dependence is given by the following general form (see, for instance, Ref. \ref{ref:Matsumoto-2017}):
\begin{align}
p^\alpha = K^\alpha_{\beta\gamma} S^\beta S^\gamma.~~~(\alpha,\beta,\gamma=x,y,z)
\label{eqn:p-general}
\end{align}
Here, $p^\alpha$ is the $\alpha$ component of the electric dipole operator.
$S^\beta$ is the $\beta$ component of the spin operator.
Since the right-hand side of Eq. (\ref{eqn:p-general}) is proportional to product of spin operators,
$K^\alpha_{\beta\gamma}$ is termed as a third-rank polar tensor.
It is real and symmetric as $K^\alpha_{\beta\gamma}=K^\alpha_{\gamma\beta}$.
This assures the Hermitian nature of the electric dipole operator.
The local symmetry of the magnetic ion is represented by the point group,
where there are various symmetry transformations.
The electric dipole in the left-hand side of Eq. (\ref{eqn:p-general}) transforms as a polar vector by the symmetry transformations,
whereas the spin operator transforms as an axial vector.
Equation (\ref{eqn:p-general}) indicates that the combination of $K^\alpha_{\beta\gamma} S^\beta S^\gamma$ must transform as a polar vector
and this restricts the possible spin dependence.
In the absence of the inversion symmetry at the magnetic ion site, $K^\alpha_{\beta\gamma}$ does not vanish.
The possible spin dependence of the electric dipole operator is then classified by the point-group symmetry.
\cite{Matsumoto-2017}
Notice that all the symmetrically-allowed spin dependences in the electric dipole can be taken into account without omission
up to the quadratic order of the spin operator.
This classification is equivalent to that for the piezoelectric tensor.
\cite{Nye-1985}
For quantum spin systems, a linear electric-field effect in paramagnetic resonance was studied,
where the electric field couples to quadrupole operators.
\cite{Mims-1976}
Since the electric field couples to an electric dipole, the quadrupole operators correspond to an electric dipole
and the classification in Ref. \ref{ref:Mims-1976} is equivalent to that in Ref. \ref{ref:Matsumoto-2017}.

For the $C_3$ symmetry, the possible spin-dependent electric dipole operator  is given by
\cite{Matsumoto-2017}
\begin{align}
\begin{pmatrix}
p^x \cr
p^y
\end{pmatrix}
&=
\begin{pmatrix}
O^{zx} & O^{x^2-y^2} & O^{yz} & O^{xy} \cr
O^{yz} & -O^{xy} & -O^{zx} & O^{x^2-y^2}
\end{pmatrix}
\begin{pmatrix}
K_1 \cr
K_2 \cr
K_1' \cr
K_2'
\end{pmatrix}, \cr
p^z &= K_3 O^{z^2}.
\label{eqn:p-C3}
\end{align}
Here, $p^x$, $p^y$, and $p^z$ are the $x$, $y$, and $z$ components of the electric dipole operator, respectively.
$K_n$ $(n=1,2,1',2',3)$ are coefficients.
In Eq. (\ref{eqn:p-C3}), quadrupole operators are defined as
\begin{align}
&O^{\alpha\beta}=S^\alpha S^\beta + S^\beta S^\alpha,~~~(\alpha,\beta=x,y,z) \cr
&O^{x^2-y^2}=(S^x)^2-(S^y)^2, \cr
&O^{z^2}=\frac{1}{\sqrt{3}}[3(S^z)^2-\bS^2].
\end{align}
Here, $S^\alpha$ is the $\alpha$ ($\alpha=x,y,z$) component of the spin operator.
Notice that the quadrupole operators vanish for $S=1/2$.

%%%%%%%%%%%%%%%%%%%%%%%%%%%%%%%%%%%%%%%%%%%%%%%%%%%%%%%%%%%%%%%%%%%%%%%%%%%%%%%%%%%%%%
\begin{figure}[t]
\begin{center}
\includegraphics[width=8cm]{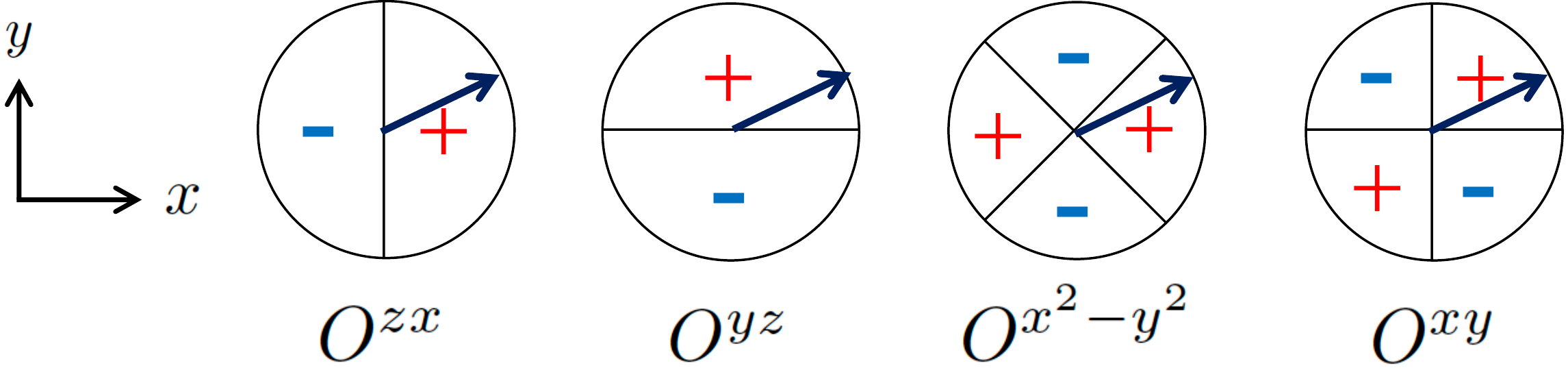}
\end{center}
\caption{
(Color online)
Schematic of sign changes in expectation values of quadrupole operators.
The solid arrow denotes the spin direction on the $xy$-plane.
The spin direction changes according to the rotation of the effective magnetic field.
The expectation value of the quadrupole operator is determined by the spin direction.
$+$ and $-$ represent the sign of the expectation value for the spin direction.
For the present spin direction, for instance, the expectation value is positive ($+$) for the all quadrupole operators.
}
\label{fig:quadrupole}
\end{figure}
%%%%%%%%%%%%%%%%%%%%%%%%%%%%%%%%%%%%%%%%%%%%%%%%%%%%%%%%%%%%%%%%%%%%%%%%%%%%%%%%%%%%%%

\begin{figure}[t]
\begin{center}
\includegraphics[width=8cm]{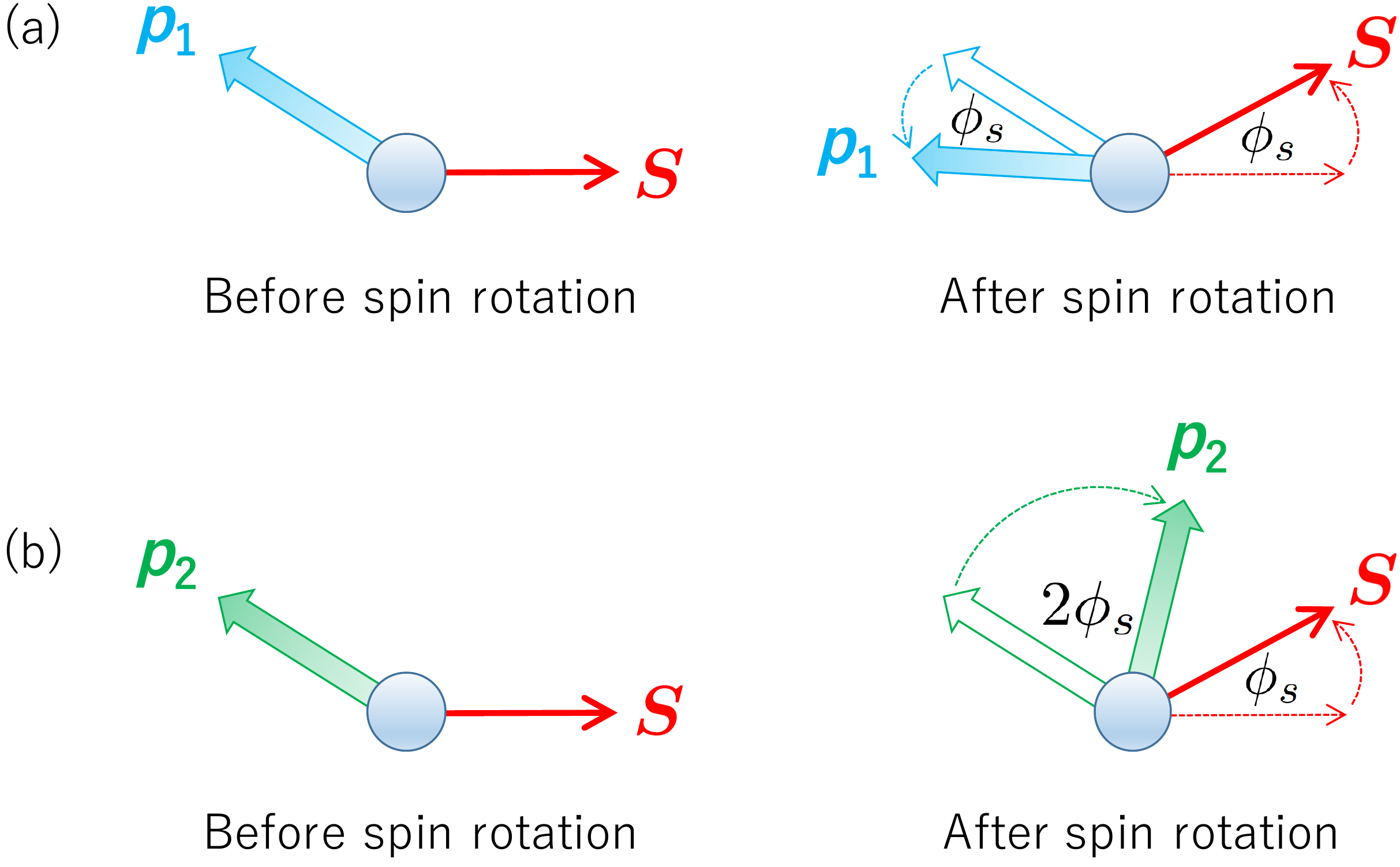}
\end{center}
\caption{
(Color online)
Schematic of rotation of electric dipole moment in the $xy$-plane.
Circles represent Co$^{2+}$ ions in $C_3$ point-group symmetry.
$\bp_1$ and $\bp_2$ represent electric dipole moments for the $\phi_s$ and $-2\phi_s$ rotation components, respectively. 
At the initial position of the spin rotation, the electric dipole moment aligns in a certain direction,
since its phase is arbitrary for the $K_1$ and $K_1'$ ($K_2$ and $K_2'$) component in $\bp_1$ ($\bp_2$), as shown in Eq. (\ref{eqn:p-C3-2}).
(a) For $\bp_1$.
After the spin rotation of $\phi_s$, $\bp_1$ rotates by $\phi_s$ in the same direction.
(b) For $\bp_2$.
After the spin rotation of $\phi_s$, $\bp_2$ rotates by $2\phi_s$ in the opposite direction.
For $\phi_s=120^\circ$, $\bp_2$ rotates by $-240^\circ=120^\circ$.
Thus, the both $\bp_1$ and $\bp_2$ are rotated by 120$^\circ$ when the spin rotates by 120$^\circ$,
indicating that the $C_3$ symmetry is satisfied.
}
\label{fig:spin-rotation}
\end{figure}

Let us consider a case where an effective magnetic field, which includes external and molecular magnetic fields, is applied in a certain direction
and suppose that the spin aligns in a ($\theta_s,\phi_s)$ direction in the polar coordinate.
Here, $\theta_s$ and $\phi_s$ are angles of the spin direction measured from the $z$- and $x$-axes, respectively.
In this case, the $\phi_s$ dependence of expectation values of the quadrupole operators are
\begin{align}
&\braket{O^{zx}} \propto \cos\phi_s,~~~
\braket{O^{yz}} \propto \sin\phi_s, \cr
&\braket{O^{xy}} \propto \sin{2\phi_s},~~~
\braket{O^{x^2-y^2}} \propto \cos{2\phi_s}, \cr
&\braket{O^{z^2}} = {\rm constant}.
\end{align}
These hold for spin $S\ge1$ systems having the quadrupole degrees of freedom, as discussed in Appendix \ref{appendix:general-S}.
Schematic of the expectation values of the quadrupole operators are shown in Fig. \ref{fig:quadrupole}.
Thus, $\braket{p^z}$ is constant for the spin rotation,
while $\braket{p^x}$ and $\braket{p^y}$ vary with $\phi_s$.
The latter components are expresses as
\begin{align}
\begin{pmatrix}
\braket{p^x} \cr
\braket{p^y}
\end{pmatrix}
&\propto
K_1
\begin{pmatrix}
\cos\phi_s \cr
\sin\phi_s
\end{pmatrix}
+ K_1'
\begin{pmatrix}
\sin\phi_s \cr
-\cos\phi_s
\end{pmatrix} \cr
&+ K_2
\begin{pmatrix}
\cos{2\phi_s} \cr
-\sin{2\phi_s}
\end{pmatrix}
+ K_2'
\begin{pmatrix}
\sin{2\phi_s} \cr
\cos{2\phi_s}
\end{pmatrix} \cr
&=
K_1
\begin{pmatrix}
\cos\phi_s \cr
\sin\phi_s
\end{pmatrix}
+ K_1'
\begin{pmatrix}
\cos(\phi_s-\frac{\pi}{2}) \cr
\sin(\phi_s-\frac{\pi}{2})
\end{pmatrix}
\label{eqn:p-C3-2} \\
&+ K_2
\begin{pmatrix}
\cos(-2\phi_s) \cr
\sin(-2\phi_s)
\end{pmatrix}
+ K_2'
\begin{pmatrix}
\cos(-2\phi_s+\frac{\pi}{2}) \cr
\sin(-2\phi_s+\frac{\pi}{2})
\end{pmatrix}.
\nonumber
\end{align}
The first two components ($K_1$ and $K_1'$) indicate that the electric dipole rotates by $\phi_s$ with the spin rotation in the $xy$-plane,
while the last two components ($K_2$ and $K_2'$) indicate that it  rotates by $-2\phi_s$, which is twice of the spin rotation in the opposite direction.
Schematic of the rotation of the electric dipole moment is shown in Fig. \ref{fig:spin-rotation}.
In the $C_3$ point group, the 120$^\circ$ rotation around the $z$-axis is the symmetry operation.
After the 120$^\circ$ spin rotation, the electric dipole must rotate by 120$^\circ$.
The component of the $\phi_s$ rotation satisfies this condition.
Since the $-2\times 120^\circ$ rotation is equivalent to the $+120^\circ$ rotation, the $-2\phi_s$ rotation component also satisfies the condition.
Thus, we emphasize that the $2\times\phi_s$ rotation component must rotate in the opposite direction relative to the spin rotation owing to the the $C_3$ symmetry.
The presence of the $-2\phi_s$ rotation component of the electric dipole is common
to the $C_3$, $D_3$, $C_{3v}$, $C_{3h}$, and $D_{3h}$ point group symmetries
which have a threefold axis along the $z$ direction.
\cite{Matsumoto-2017}

%%%%%%%%%%%%%%%%%%%%%%%%%%%%%%%%%%%%%%%%%%%%%%%%%%%%%%%%%%%%%%%%%%%%%%%%%%%%%%%%%%%%%%
%%%%%%%%%%%%%%%%%%%%%%%%%%%%%%%%%%%%%%%%%%%%%%%%%%%%%%%%%%%%%%%%%%%%%%%%%%%%%%%%%%%%%%

In cubic systems, there are also threefold axes along the [111] and its equivalent directions.
In this case, we can expect the same electromagnetic properties under the broken inversion symmetry.
$T$ and $T_d$ are the point group symmetries that match the conditions.
The possible spin-dependent electric dipole operators for $T$ and $T_d$ are listed in Ref. \ref{ref:Matsumoto-2017}.
In Appendix \ref{appendix:cubic}, we take the $\tz$-axis along the [111] direction,
whereas the $\tx$- and $\ty$-axes are taken perpendicular to the [111] direction (see Fig. \ref{fig:cubic}).
The electric dipole in the $\tx\ty\tz$ coordinate is given by Eq. (\ref{eqn:p-cubic}).
Comparing Eqs. (\ref{eqn:p-C3}) and (\ref{eqn:p-cubic}),
we find the following correspondence between the coupling constants:
$(K_1,K_2,K_1',K_2',K_3)\leftrightarrow(-\frac{1}{\sqrt{3}}K,\sqrt{\frac{2}{3}}K,0,0,K)$.
This indicates that the both $\phi_s$ and $-2\phi_s$ rotation components of the electric dipole exist in the cubic systems.
In addition, notice that the ratio of the couplings for the two components ($K_1/K_2$) is fixed as the consequence of the group-theoretical analysis.
These points can be checked by experiments in the cubic ($T$ and $T_d$) systems.

Thus, it is the universal property that the electric dipole rotates with the rotation of the spin around the threefold axis
when a magnetic ion ($S\ge 1$) occupies a site lacking the inversion symmetry.
There are two components.
One rotates in the same direction of the spin, while the other rotates in the opposite direction at the twice speed.

%%%%%%%%%%%%%%%%%%%%%%%%%%%%%%%%%%%%%%%%%%%%%%%%%%%%%%%%%%%%%%%%%%%%%%%%%%%%%%%%%%%%%%%%%%%%%%%%%%%
\subsection{Spin-dependent electric dipole in \Co}
%%%%%%%%%%%%%%%%%%%%%%%%%%%%%%%%%%%%%%%%%%%%%%%%%%%%%%%%%%%%%%%%%%%%%%%%%%%%%%%%%%%%%%%%%%%%%%%%%%%

In \Co, the honeycomb structure consists of non-equivalent Co(1) and Co(2) sites, as shown in Fig. \ref{fig:crystal}.
We focus on the Co(1) site here, since the same discussion holds for the Co(2) site.
The four equivalent sites for the Co(1) are related by the inversion center and the twofold axis.
This holds for the spin-dependences of the electric dipoles of those sites.
For the site-index of Co(1) shown in Fig. \ref{fig:crystal}(b),
we use $(a,b,a',b')$ here instead of $(a_1, b_1, a'_1, b'_1)$ for simplicity.

As in Eq. (\ref{eqn:p-C3}), the spin-dependent electric dipole operator on the $a$ site is given by
\begin{align}
\begin{pmatrix}
p_a^x \cr
p_a^y
\end{pmatrix}
&=
\begin{pmatrix}
O_a^{zx} & O_a^{x^2-y^2} & O_a^{yz} & O_a^{xy} \cr
O_a^{yz} & -O_a^{xy} & -O_a^{zx} & O_a^{x^2-y^2}
\end{pmatrix}
\begin{pmatrix}
K_1 \cr
K_2 \cr
K_1' \cr
K_2'
\end{pmatrix}, \cr
p_a^z &= O_a^{z^2} K_3.
\label{eqn:p-A}
\end{align}
Here, $(p_a^x,p_a^y,p_a^z)$ is the electric dipole at the $a$ site.
$O_a^{\alpha\beta}$ ($\alpha=x,y,z$) and $O_a^{x^2-y^2}$ are quadrupole operators at the $a$ site.

In the \Pc~space group, there is a inversion center between the $a$ and $b$ sites [see Fig. \ref{fig:crystal}(b)].
The inversion operation transforms $(x,y,z)\rightarrow(-x,-y,-z)$.
The electric dipole operator is transformed as $(p_a^x,p_a^y,p_a^z)\rightarrow(-p_b^x,-p_b^y,-p_b^z)$.
The spin operators are transformed as $(S_a^x,S_a^y,S_a^z)\rightarrow (S_b^x,S_b^y,S_b^z)$.
The quadrupole operators are transformed as $O_a^m \rightarrow O_b^m$ $(m=zx,xy,yz,x^2-y^2,z^2)$.
The electric dipole at the $b$ site is then expressed with the coefficients for the $a$ site as
\begin{align}
\begin{pmatrix}
p_b^x \cr
p_b^y
\end{pmatrix}
&=
\begin{pmatrix}
O_b^{zx} & O_b^{x^2-y^2} & O_b^{yz} & O_b^{xy} \cr
O_b^{yz} & -O_b^{xy} & -O_b^{zx} & O_b^{x^2-y^2}
\end{pmatrix}
\begin{pmatrix}
-K_1 \cr
-K_2 \cr
-K_1' \cr
-K_2'
\end{pmatrix}, \cr
p_b^z &= - O_b^{z^2} K_3.
\label{eqn:pB-inverse}
\end{align}

Next, we consider the twofold axis.
The $a$ and $b'$ sites are related by the twofold axis, as shown in Fig. \ref{fig:crystal}(b).
The $\pi$ rotation along the $y$-axis transforms $(x,y,z)\rightarrow(-x,y,-z)$.
The electric dipole operator is transformed as $(p_a^x,p_a^y,p_a^z)\rightarrow(-p_{b'}^x,p_{b'}^y,-p_{b'}^z)$.
The spin operators are transformed as $(S_a^x,S_a^y,S_a^z)\rightarrow (-S_{b'}^x,S_{b'}^y,-S_{b'}^z)$.
The quadrupole operators are transformed as
$(O_a^{zx},O_a^{xy},O_a^{yz},O_a^{x^2-y^2},O_a^{z^2}) \rightarrow (O_{b'}^{zx},-O_{b'}^{xy},-O_{b'}^{yz},O_{b'}^{x^2-y^2},O_{b'}^{z^2})$.
The electric dipole at the $b'$ site is then expressed as
\begin{align}
\begin{pmatrix}
p_{b'}^x \cr
p_{b'}^y
\end{pmatrix}
&=
\begin{pmatrix}
O_{b'}^{zx} & O_{b'}^{x^2-y^2} & O_{b'}^{yz} & O_{b'}^{xy} \cr
O_{b'}^{yz} & -O_{b'}^{xy} & -O_{b'}^{zx} & O_{b'}^{x^2-y^2}
\end{pmatrix}
\begin{pmatrix}
-K_1 \cr
-K_2 \cr
 K_1' \cr
 K_2'
\end{pmatrix}, \cr
p_{b'}^z &= - O_{b'}^{z^2} K_3.
\label{eqn:pB-2-fold}
\end{align}

We consider next the $a'$ site.
The $a'$ and $b'$ sites are related by the inversion center.
The electric dipole at the $a'$ site is then expressed as
\begin{align}
\begin{pmatrix}
p_{a'}^x \cr
p_{a'}^y
\end{pmatrix}
&=
\begin{pmatrix}
O_{a'}^{zx} & O_{a'}^{x^2-y^2} & O_{a'}^{yz} & O_{a'}^{xy} \cr
O_{a'}^{yz} & -O_{a'}^{xy} & -O_{a'}^{zx} & O_{a'}^{x^2-y^2}
\end{pmatrix}
\begin{pmatrix}
K_1 \cr
K_2 \cr
-K_1' \cr
-K_2'
\end{pmatrix}, \cr
p_{a'}^z &= O_{a'}^{z^2} K_3.
\label{eqn:pA'}
\end{align}
Notice that Eq. (\ref{eqn:pA'}) can be obtained from Eq. (\ref{eqn:pB-inverse}) by the transformation of the twofold axis.
This means that the obtained relations
in Eqs. (\ref{eqn:p-A}), (\ref{eqn:pB-inverse}), (\ref{eqn:pB-2-fold}), and (\ref{eqn:pA'}) are consistent.
In addition to the inversion and twofold rotation symmetries, there is $c$-glide symmetry in the \Pc~space group.
\cite{IN-table}
Since the $c$-glide can be reproduced by the combination of the inversion and twofold transformations,
\cite{note-c-glide}
it does not alter the result of Eqs. (\ref{eqn:p-A}), (\ref{eqn:pB-inverse}), (\ref{eqn:pB-2-fold}), and (\ref{eqn:pA'}).

Finally, we discuss the expectation value of the total electric dipole.
Since the magnetic structure is uniform along the $c$-axis,
it is the same for the A ($a,a'$) and B ($b,b'$) sites (see Fig. \ref{fig:crystal}).
The expectation values of the quadrupole operators satisfy
$\braket{O^m_a}=\braket{O^m_{a'}}\equiv\braket{O^m_{\rA}}$ and $\braket{O^m_b}=\braket{O^m_{b'}}\equiv\braket{O^m_{\rB}}$.
After adding the contributions from the four sites, we obtain the following total electric dipole per a unit cell (electric polarization) for the Co(1) site:
\begin{align}
&
\begin{pmatrix}
\braket{P^x} \cr
\braket{P^y}
\end{pmatrix}
=
\begin{pmatrix}
\braket{p_a^x + p_{a'}^x + p_b^x + p_{b'}^x} \cr
\braket{p_a^y + p_{a'}^y + p_b^y + p_{b'}^y}
\end{pmatrix} \cr
&=2
\begin{pmatrix}
\braket{O_\rA^{zx}}-\braket{O_\rB^{zx}} & \braket{O_\rA^{x^2-y^2}}-\braket{O_\rB^{x^2-y^2}} \cr
\braket{O_\rA^{yz}}-\braket{O_\rB^{yz}} & -\left(\braket{O_\rA^{xy}}-\braket{O_\rB^{xy}}\right)
\end{pmatrix}
\begin{pmatrix}
K_1 \cr
K_2
\end{pmatrix}, \cr
&\braket{P^z} = \braket{p_a^z + p_{a'}^z + p_b^z + p_{b'}^z} \cr
&~~~~~~
= 2 \left( \braket{O_\rA^{z^2}} - \braket{O_\rB^{z^2}} \right) K_3.
\label{eqn:p-total}
\end{align}
Here, the $K_1'$ and $K_2'$ components vanish by cancellation of the A and B sites.

The same result can be obtained for the Co(2) site,
although the coupling constants $(K_1,K_2,K_3)$ may be different from those of the Co(1) site.
When we add the contributions from the Co(1) and Co(2) sites, the coupling constants are modified as
$\tK_n=K_n(1)+K_n(2)$ ($n=1,2,3$).
Here, $K_n(1)$ is for the Co(1) site, whereas $K_n(2)$ is for the Co(2) site.
Thus, Eq. (\ref{eqn:p-total}) provides the possible spin dependence of the electric polarization in \Co.
In the next subsection, we explain the observed electromagnetic effect.

%%%%%%%%%%%%%%%%%%%%%%%%%%%%%%%%%%%%%%%%%%%%%%%%%%%%%%%%%%%%%%%%%%%%%%%%%%%%%%%%%%%%%%%%%%%%%%%%%%%
\subsection{Magnetic field-induced electric dipole in \Co}
%%%%%%%%%%%%%%%%%%%%%%%%%%%%%%%%%%%%%%%%%%%%%%%%%%%%%%%%%%%%%%%%%%%%%%%%%%%%%%%%%%%%%%%%%%%%%%%%%%%

%%%%%%%%%%%%%%%%%%%%%%%%%%%%%%%%%%%%%%%%%%%%%%%%%%%%%%%%%%%%%%%%%%%%%%%%%%%%%%%%%%%%%%
\begin{figure}[t]
\begin{center}
\includegraphics[width=5.5cm]{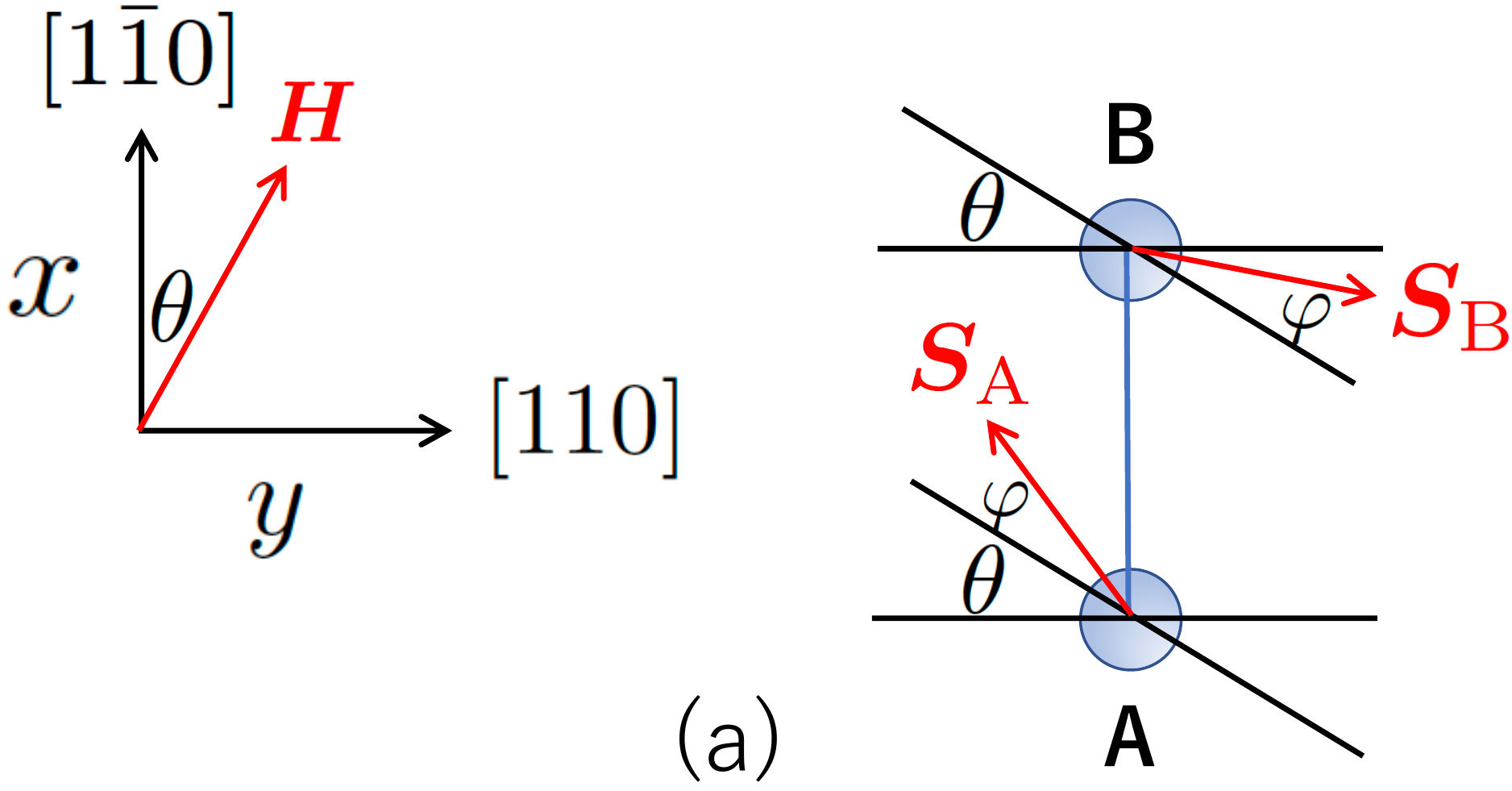}
\includegraphics[width=5.5cm]{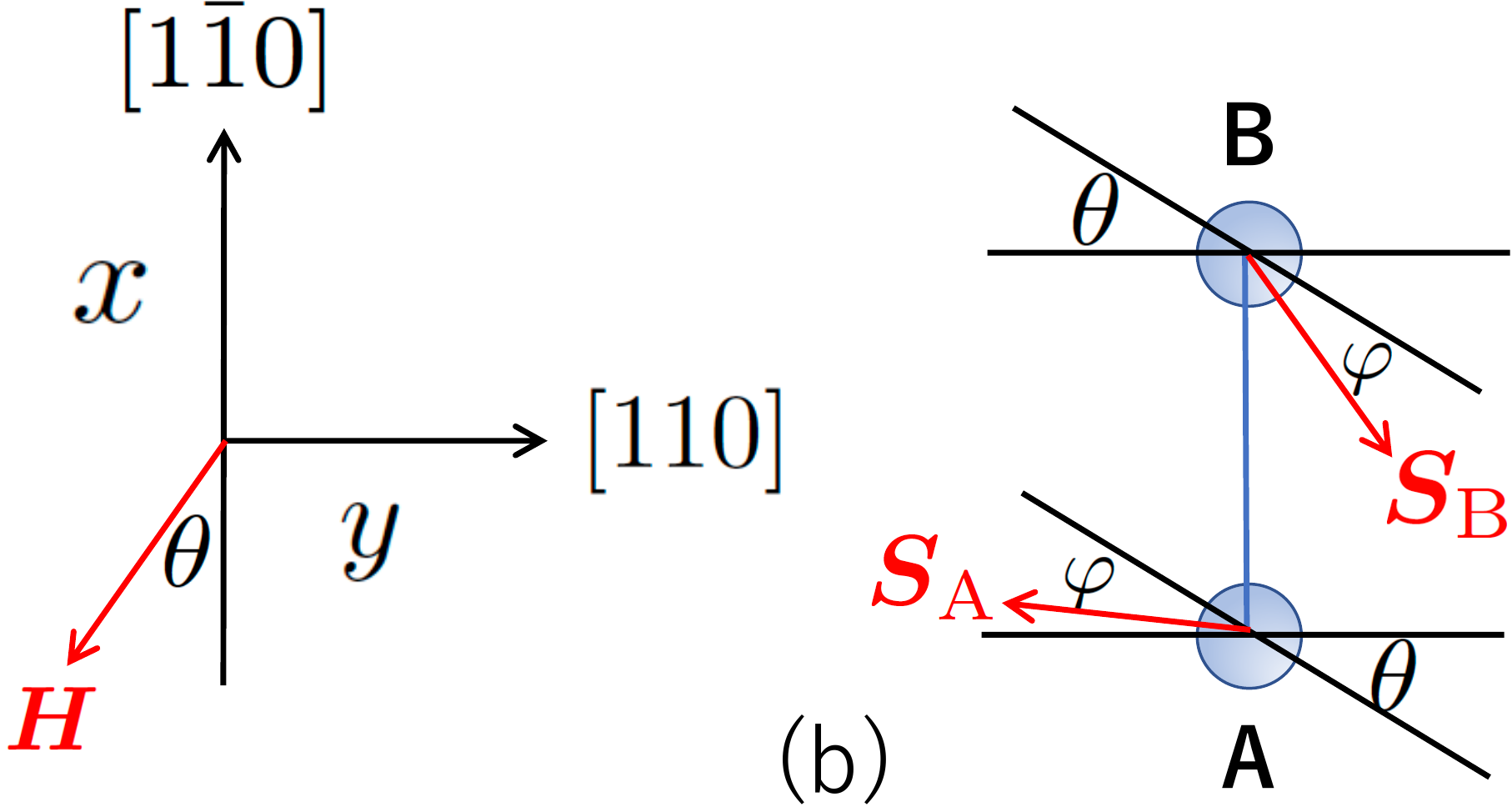}
\end{center}
\caption{
(Color online)
Schematic of the magnetic structure under a finite magnetic field applied in the $ab$-plane.
A and B represent the two Co sites.
$\bS_\rA$ and $\bS_\rB$ represent the magnetic moment at the A and B sites, respectively.
(a) For field-rotating process.
$\theta$ is the angle of the field measured from the $x$-axis.
The AF moment tends to align perpendicular to the field $\bH$.
$\varphi$ is the canting angle of the magnetic moment.
A net magnetic moment is induced by the canting.
(b) For field-sweeping process.
The direction of the magnetic field is reversed.
}
\label{fig:field}
\end{figure}
%%%%%%%%%%%%%%%%%%%%%%%%%%%%%%%%%%%%%%%%%%%%%%%%%%%%%%%%%%%%%%%%%%%%%%%%%%%%%%%%%%%%%%

\Co~shows an AF long-range order below the N\'{e}el temperature.
The ordered moment aligns in the $ab$-plane owing to the strong easy-plane single-ion anisotropy.
\cite{Khanh-2016,Khanh-2017,Deng-2018}
Under a finite magnetic field applied in the $ab$-plane, the AF moment shows the tendency to align perpendicular to the field in the plane.
The schematic of the magnetic moment under the field is shown in Fig. \ref{fig:field}(a).
Expectation values of the spin operators at the A and B sites can be expressed as
\begin{align}
\begin{pmatrix}
\braket{S_\rA^x} \cr
\braket{S_\rA^y} \cr
\braket{S_\rA^z}
\end{pmatrix}
&=
\begin{pmatrix}
  S_\perp \sin(\theta+\varphi) \cr
- S_\perp \cos(\theta+\varphi) \cr
  S_\parallel
\end{pmatrix}, \cr
\begin{pmatrix}
\braket{S_\rB^x} \cr
\braket{S_\rB^y} \cr
\braket{S_\rB^z}
\end{pmatrix}
&=
\begin{pmatrix}
- S_\perp \sin(\theta-\varphi) \cr
  S_\perp \cos(\theta-\varphi) \cr
  S_\parallel
\end{pmatrix}.
\label{eqn:S-expect}
\end{align}
Following Ref. \ref{ref:Khanh-2017}, we take $\theta$ as the angle of the external magnetic field in the $ab$-plane (see Fig. \ref{fig:field}).
$\varphi$ is a canting angle of the magnetic moment.
It can be determined by a mean-field theory under the external field.
$S_\perp$ and $S_\parallel$ are constants for the $ab$-plane and the $z$ components of the expectation values, respectively.
We assume that the external magnetic field has a $z$ component and there is a finite moment along the $z$-axis, i.e. $S_\parallel\neq 0$.
Expectation values of the quadrupole operators at the A and B sites can be expressed as
\begin{align}
\begin{pmatrix}
\braket{O_\rA^{zx}} \cr
\braket{O_\rA^{yz}} \cr
\braket{O_\rA^{x^2-y^2}} \cr
\braket{O_\rA^{xy}} \cr
\braket{O_\rA^{z^2}}
\end{pmatrix}
&=
\begin{pmatrix}
  O_1 \sin{(\theta+\varphi)} \cr
- O_1 \cos{(\theta+\varphi)} \cr
- O_2 \cos{2(\theta+\varphi)} \cr
- O_2 \sin{2(\theta+\varphi)} \cr
  O_z \cr
\end{pmatrix}, \cr
\begin{pmatrix}
\braket{O_\rB^{zx}} \cr
\braket{O_\rB^{yz}} \cr
\braket{O_\rB^{x^2-y^2}} \cr
\braket{O_\rB^{xy}} \cr
\braket{O_\rB^{z^2}}
\end{pmatrix}
&=
\begin{pmatrix}
- O_1 \sin{(\theta-\varphi)} \cr
  O_1 \cos{(\theta-\varphi)} \cr
- O_2 \cos{2(\theta-\varphi)} \cr
- O_2 \sin{2(\theta-\varphi)} \cr
  O_z \cr
\end{pmatrix}.
\label{eqn:O-expect}
\end{align}
Here, $O_1$, $O_2$ and $O_z$ are constants which can be determined by the easy-plane anisotropy and the effective magnetic field.
Now, the $\theta$ dependence of the expectation values of the quadrupole operators are expressed
by the canting angle $\varphi$ and the three constants ($O_1$, $O_2$, and $O_z$).
After substituting Eq. (\ref{eqn:O-expect}) into Eq. (\ref{eqn:p-total}), we obtain
\begin{align}
\begin{pmatrix}
\braket{P^x} \cr
\braket{P^y}
\end{pmatrix}
&=
4\tK_1 O_1 \cos\varphi
\begin{pmatrix}
  \sin{\theta} \cr
- \cos{\theta}
\end{pmatrix} \cr
&+ 4\tK_2 O_2 \sin{2\varphi}
\begin{pmatrix}
\sin{2\theta} \cr
\cos{2\theta}
\end{pmatrix}, \label{eqn:p-result} \\
\braket{P^z} &= 0.
\nonumber
\end{align}
Here, $\tK_1$ and $\tK_2$ are coupling constants for the total electric polarization [see the discussion below Eq. (\ref{eqn:p-total})].
This indicates that no electric polarization is induced in the $z$ direction.
It is owing to the inversion center (or twofold axis) located at the center of the A and B Co sites.
At each Co site, the spin-dependent electric dipole operator can exist in the from given by Eq. (\ref{eqn:p-C3}),
however, Eq. (\ref{eqn:p-result}) indicates that the $K_1'$, $K_2'$, and $K_3$ components are not active in the electric polarization.

%%%%%%%%%%%%%%%%%%%%%%%%%%%%%%%%%%%%%%%%%%%%%%%%%%%%%%%%%%%%%%%%%%%%%%%%%%%%%%%%%%%%%%
\begin{figure}[t]
\begin{center}
\includegraphics[width=8cm]{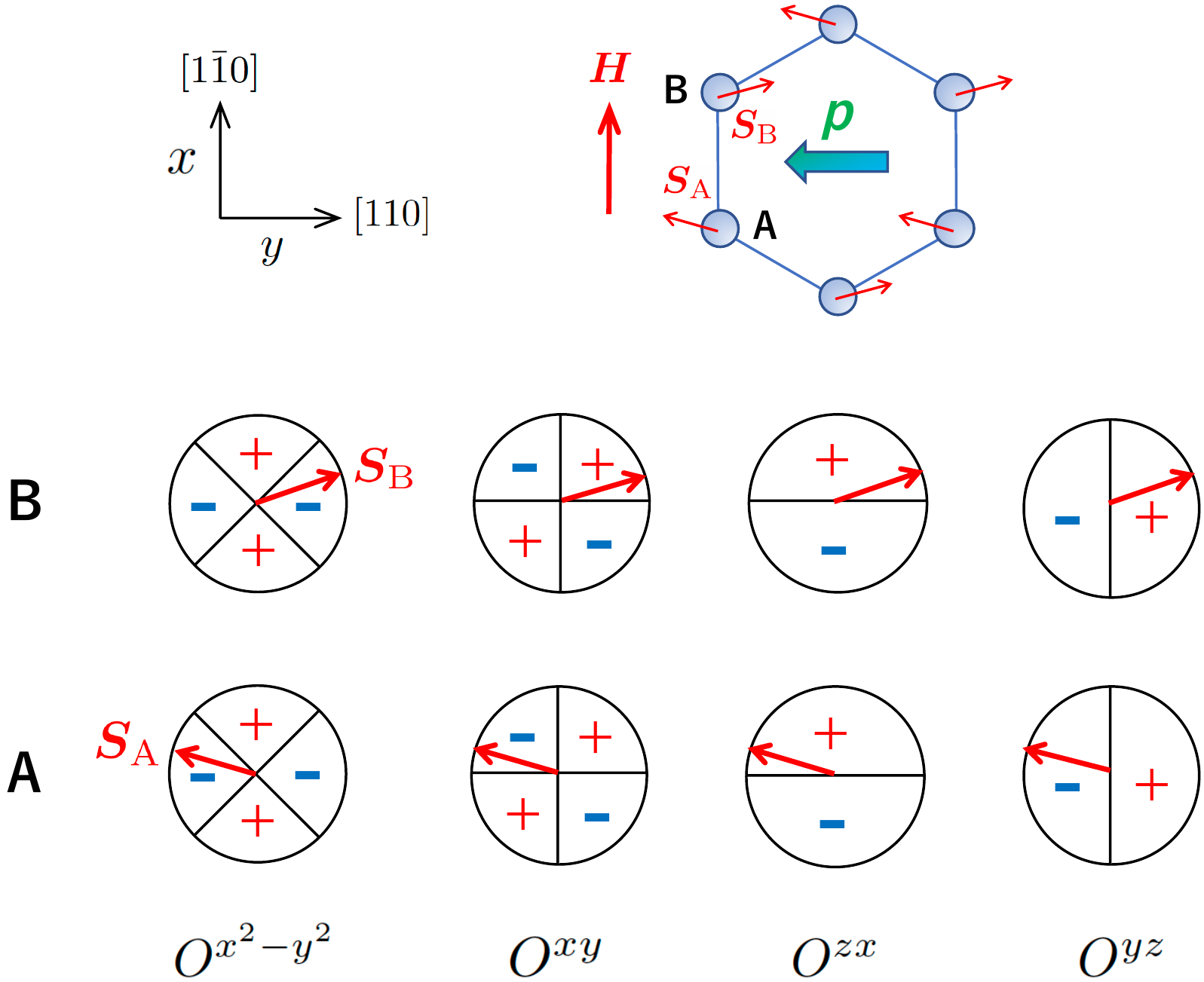}
\end{center}
\caption{
(Color online)
Schematics of spin configuration on the honeycomb lattice
and sign changes in expectation values of quadrupole operators for $\bH\parallel [1\bar{1}0]$ ($\theta=0$).
Notice that the $x$ and $y$ coordinates are taken differently from Fig. \ref{fig:quadrupole}.
$\bS_\rA$ and $\bS_\rB$ represent the spin direction at the A and B sites, respectively.
The spin direction changes with the rotation of the external magnetic field
and the expectation values of the quadrupole operators change accordingly.
This determines the electric polarization through Eq. (\ref{eqn:p-total}).
}
\label{fig:pxy-h=0}
\end{figure}
%%%%%%%%%%%%%%%%%%%%%%%%%%%%%%%%%%%%%%%%%%%%%%%%%%%%%%%%%%%%%%%%%%%%%%%%%%%%%%%%%%%%%%

Under the magnetic field $\bH\parallel [1\bar{1}0]$, for instance,
we show schematics of spin configuration and expectation values of quadrupole operators in Fig. \ref{fig:pxy-h=0}.
We can see that
$\braket{O^{zx}_\rA}=\braket{O^{zx}_\rB}$, $\braket{O^{x^2-y^2}_\rA}=\braket{O^{x^2-y^2}_\rB}$,
$\braket{O^{xy}_\rA}=-\braket{O^{xy}_\rB}$, and $\braket{O^{yz}_\rA}=-\braket{O^{yz}_\rB}$.
From Eq. (\ref{eqn:p-total}), these indicate that $\braket{p^x}=0$ and $\braket{p^y}\neq 0$
and the electric polarization appears in the $[110]$ direction for $\bH\parallel [1\bar{1}0]$ (see Fig. \ref{fig:pxy-h=0}).

In the absence of the external magnetic field $(\bH=0)$,
the magnetic structure is collinear $(\varphi=0)$ and $O_1=0$ in Eq. (\ref{eqn:p-result}).
Therefore, the polarization does no appear.
From a symmetry point of view, this can be understood as follows.
The crystal structure has the inversion symmetry, as shown in Fig. \ref{fig:crystal}.
Below the N\'{e}el temperature, the magnetic structure breaks the inversion symmetry.
However, the AF structure for $\bH=0$ possesses an $I \Theta$ symmetry transformation.
Here, $I$ and $\Theta$ are inversion and time-reversal operations, respectively.
The ground state $|g\rangle$ is then an eigenstate of $I \Theta$ as $I\Theta |g\rangle=\lambda |g\rangle$ with $|\lambda|^2=1$.
In this case, the expectation value of electric polarization $\bP$ is calculated as
\cite{J-J-Sakurai-2017}
\begin{align}
\braket{g| \bP |g}
&= \braket{\Theta g| \Theta \bP \Theta^{-1} | \Theta g} \cr
&= \braket{I \Theta g| I \bP I^{-1} | I \Theta g} \cr
&= - \braket{g| \bP |g}.
\label{eqn:P-AF-0}
\end{align}
Here, we used $\Theta \bP \Theta^{-1}=\bP$, $I \bP I^{-1}=-\bP$, and $|\lambda|^2=1$.
Equation (\ref{eqn:P-AF-0}) indicates that $\braket{g|\bP|g}=0$ owing to the $I\Theta$ symmetry transformation.
Thus, the electric polarization disappears in the absence of the external magnetic field.
Under a finite magnetic field, $I\Theta$ is not a symmetry transformation any more and the polarization can be finite.

%%%%%%%%%%%%%%%%%%%%%%%%%%%%%%%%%%%%%%%%%%%%%%%%%%%%%%%%%%%%%%%%%%%%%%%%%%%%%%%%%%%%%%%%%%%%%%%%%%%
\subsubsection{$2\theta$-rotation component}
%%%%%%%%%%%%%%%%%%%%%%%%%%%%%%%%%%%%%%%%%%%%%%%%%%%%%%%%%%%%%%%%%%%%%%%%%%%%%%%%%%%%%%%%%%%%%%%%%%%

%%%%%%%%%%%%%%%%%%%%%%%%%%%%%%%%%%%%%%%%%%%%%%%%%%%%%%%%%%%%%%%%%%%%%%%%%%%%%%%%%%%%%%
\begin{figure}[t]
\begin{center}
\includegraphics[width=7cm]{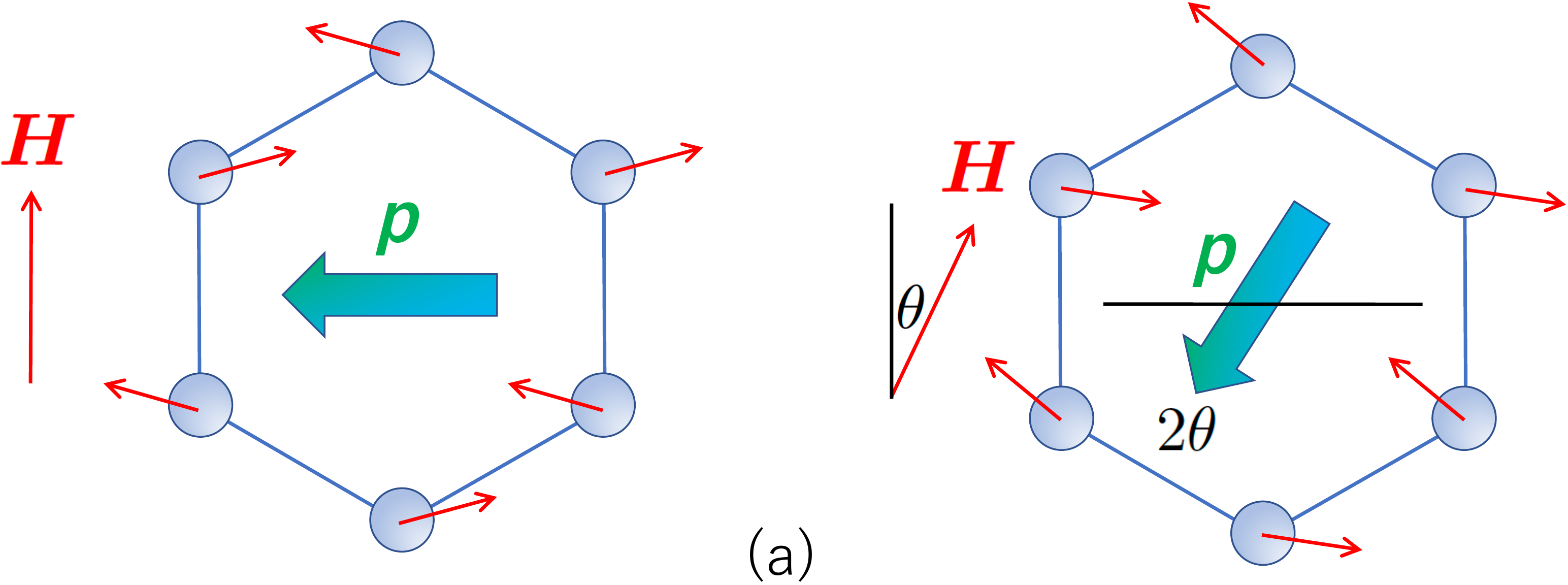}
\includegraphics[width=7cm]{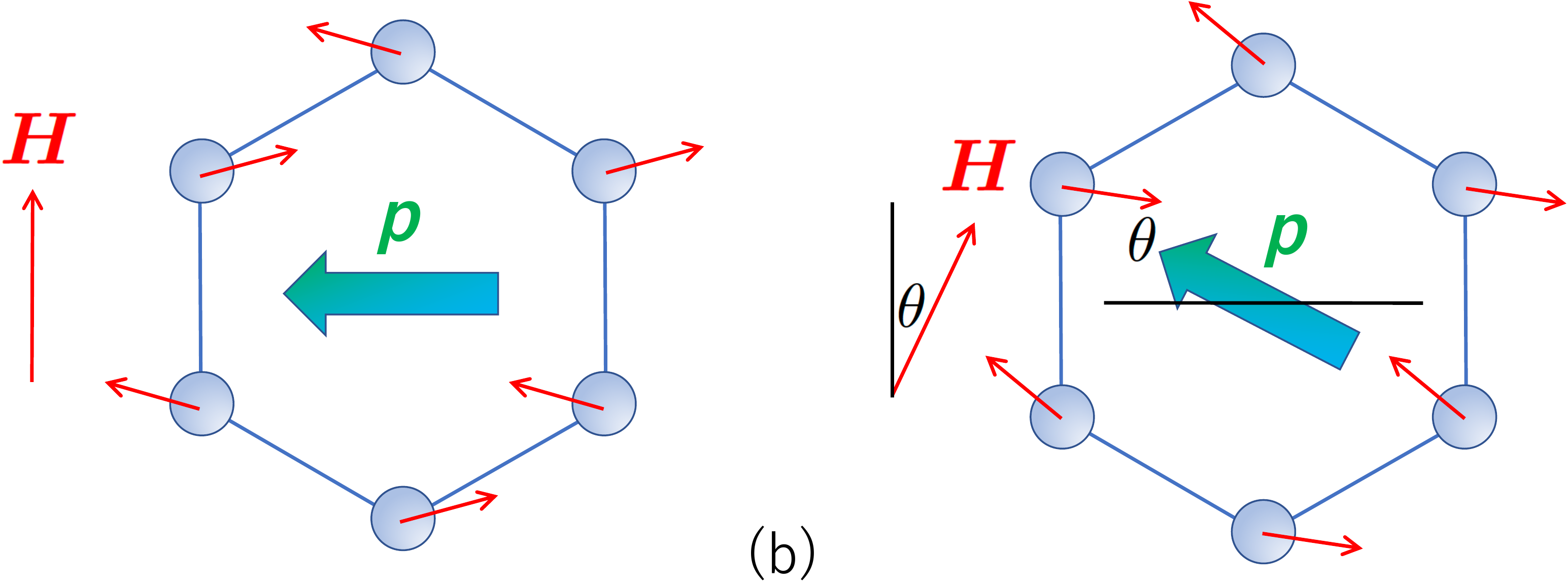}
\end{center}
\caption{
(Color online)
Schematic of rotation of the electric polarization with the rotation of the external magnetic field.
(a) $2\theta$-rotation component.
The direction of the magnetic field rotates by $\theta$ in the clockwise direction in the $ab$-plane.
The electric polarization rotates by $2\theta$ in the counterclockwise direction.
This picture is essentially the same as Fig. 4(a) in Ref. \ref{ref:Khanh-2017}.
Here, we assumed a negative coupling constant ($\tK_2<0$) to compare the result with the experiment.
\cite{Khanh-2017}
(b) $\theta$-rotation component.
In this case, it is assumed that the magnetic field has a $z$ component as well.
The electric polarization rotates by $\theta$ in the clockwise direction.
}
\label{fig:rotation}
\end{figure}
%%%%%%%%%%%%%%%%%%%%%%%%%%%%%%%%%%%%%%%%%%%%%%%%%%%%%%%%%%%%%%%%%%%%%%%%%%%%%%%%%%%%%%

As revealed by Khanh and coworkers, the electric polarization appears in the $ab$-plane
when an external magnetic field is applied in the $ab$-plane.
The direction of the polarization rotates by $2\theta$ in the opposite direction relative to the $\theta$ rotation of the magnetic field.
\cite{Khanh-2016,Khanh-2017}
This property can be understood by the second component proportional to $\tK_2$ in Eq. (\ref{eqn:p-result}).
The induced electric polarization shows the $2\theta$ rotation in the opposite direction of that of the external magnetic field,
as shown in Fig. \ref{fig:rotation}(a).
When the field is applied in the $ab$-plane, no magnetic moment is induced in the $z$ direction
and the expectation values of the quadrupole operators are $\braket{O^{zx}}=\braket{O^{yz}}=0$, i.e. $O_1=0$ in Eq. (\ref{eqn:p-result}).
Therefore, only the $2\theta$-rotation component remains and this is in agreement with the experimental result.

Another interesting experimental result is that the electric polarization
shows a linear magnetic-field dependence and changes its sign when the direction of the magnetic field is reversed (field-sweeping process).
\cite{Khanh-2016,Khanh-2017}
This is also explained by the $2\theta$-rotation component in Eq. (\ref{eqn:p-result}).
When the magnetic field is reversed, the magnetic moments align as shown in Fig. \ref{fig:field}(b).
This indicates that the canting angle is reversed as $\varphi\rightarrow -\varphi$.
The coefficient $\sin{2\varphi}$ for the $2\theta$-rotation component in Eq. (\ref{eqn:p-result})
changes as $\sin{2\varphi} \rightarrow -\sin{2\varphi}$.
For a weak external magnetic field $H$, the canting angle $\varphi$ is small and $\sin{2\varphi}\propto H$ is expected.
The value of $O_2$ in Eq. (\ref{eqn:p-result}) is determined by the molecular field and is almost independent of $H$.
Thus, the magnitude of the $2\theta$-rotation component is proportional to $H$
and we can understand the linear electromagnetic effect of the field-sweeping process.
Although the directions of the magnetic fields are the same for both field-rotating and field-sweeping processes,
the spin configurations are different as shown in Fig. \ref{fig:field}.
This is the reason why the different response to the external magnetic field is observed in the induced electric polarization.

%%%%%%%%%%%%%%%%%%%%%%%%%%%%%%%%%%%%%%%%%%%%%%%%%%%%%%%%%%%%%%%%%%%%%%%%%%%%%%%%%%%%%%%%%%%%%%%%%%%
\subsubsection{$\theta$-rotation component}
%%%%%%%%%%%%%%%%%%%%%%%%%%%%%%%%%%%%%%%%%%%%%%%%%%%%%%%%%%%%%%%%%%%%%%%%%%%%%%%%%%%%%%%%%%%%%%%%%%%

The first component proportional to $\tK_1$ in Eq. (\ref{eqn:p-result}) represents an electric polarization
whose direction rotates by $\theta$ in the same direction of the external magnetic field
[see Fig. \ref{fig:rotation}(b)].
The $\theta$ rotation originates from the $O^{zx}$ and $O^{yz}$ quadrupoles, as shown in Eq. (\ref{eqn:O-expect}).
This component can be present when the external magnetic field inclines toward the $z$ direction from the $ab$-plane.
When the applied magnetic field has a finite $z$ component, i.e. $H_z\neq 0$,
we can expect finite values of $S_\parallel$ and $O_1$ in Eqs. (\ref{eqn:S-expect}) and (\ref{eqn:O-expect}), respectively.
In this case, the $\theta$-rotation component in Eq. (\ref{eqn:p-result}) emerges in addition to the $2\theta$-rotation one.
The characteristic point of the emergent component is that the electric polarization
rotates in the same direction relative to the rotation of the external magnetic field,
i.e., it rotates by $\theta$ in the opposite direction relative to the $2\theta$-rotation component (see Fig. \ref{fig:rotation}).
The magnitude of the $\theta$-rotation component is proportional to $H_z$ for weak $H_z$,
since $O_1\propto H_z$ and $\cos\varphi\simeq 1$ in Eq. (\ref{eqn:p-result}).

%%%%%%%%%%%%%%%%%%%%%%%%%%%%%%%%%%%%%%%%%%%%%%%%%%%%%%%%%%%%%%%%%%%%%%%%%%%%%%%%%%%%%%
\begin{figure}[t]
\begin{center}
\includegraphics[width=5.5cm]{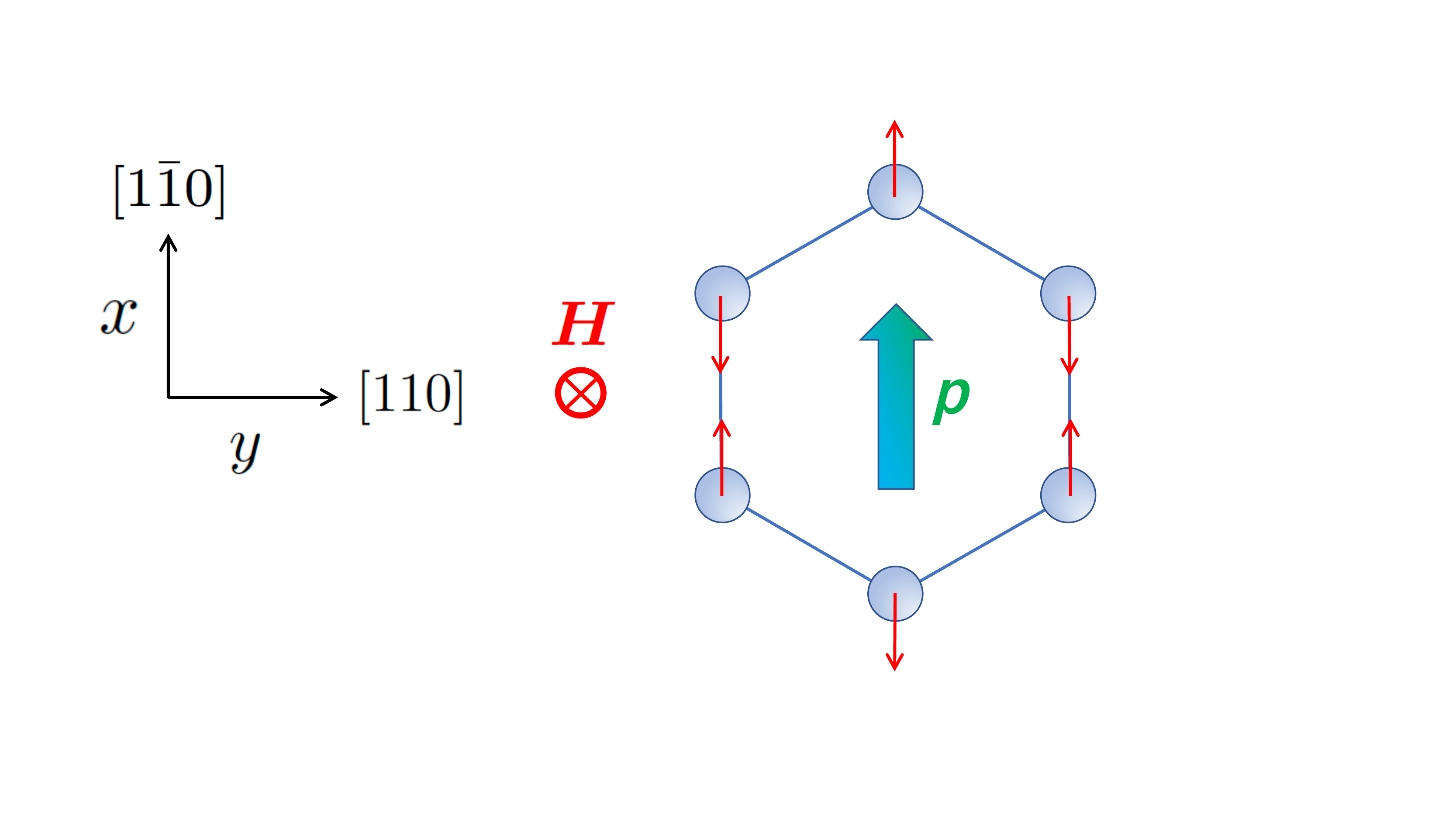}
\end{center}
\caption{
(Color online)
Schematic of the magnetic structure under the magnetic field in the $z$ direction.
The AF moment aligns in the $x$ direction, which is parallel to the easy-axis ($[1\bar{1}0]$ direction).
The electric polarization $\bP$ induced in the $x$ direction is owing to a finite uniform magnetic moment in the $z$ direction.
}
\label{fig:Hz}
\end{figure}
%%%%%%%%%%%%%%%%%%%%%%%%%%%%%%%%%%%%%%%%%%%%%%%%%%%%%%%%%%%%%%%%%%%%%%%%%%%%%%%%%%%%%%

In the absence of the external magnetic field, it is reported that the ordered moment aligns in the $[1\bar{1}0]$ direction.
\cite{Khanh-2016,Khanh-2017,Deng-2018}
This indicates that there is a weak easy-axis anisotropy in the $ab$-plane.
It appears as the spin-flop transition when the magnetic field is applied in the $[1\bar{1}0]$ direction.
\cite{Kolodiazhnyi-2011,Cao-2015,Khanh-2016,Khanh-2017}
Let us discuss a case when a magnetic field is applied along the $z$ direction,
where the AF moment aligns in the $[1\bar{1}0]$ direction with a finite uniform moment in the $z$ direction.
As shown in Fig. \ref{fig:Hz}, this magnetic structure corresponds to the case
of $\theta=\pi$ and $\varphi=0$ in Eq. (\ref{eqn:p-result}).
In this case, the $2\theta$-rotation component is not induced in the absence of the canting angle ($\varphi=0$).
In contrast to this, the expectation value of the $O^{zx}$ quadrupole can be finite.
For a weak magnetic field, $O_1\propto H$ and the electric polarization appears
in the $x$ direction as $\braket{p^x}\propto H$.
It is worthwhile to explore the above properties of the $\theta$-rotation component in \Co~by future experiments.

%%%%%%%%%%%%%%%%%%%%%%%%%%%%%%%%%%%%%%%%%%%%%%%%%%%%%%%%%%%%%%%%%%%%%%%%%%%%%%%%%%%%%%%%%%%%%%%%%%%
\section{Optical Properties}
%%%%%%%%%%%%%%%%%%%%%%%%%%%%%%%%%%%%%%%%%%%%%%%%%%%%%%%%%%%%%%%%%%%%%%%%%%%%%%%%%%%%%%%%%%%%%%%%%%%

Optical properties are also unique in the presence of magnetoelectric effects.
In this section, we theoretically study possible optical absorption in \Co,
where the electric field component of light can participate in the excitation process as well as the magnetic component.
This can appear as various types of dichroism, as in other multiferroic materials.

%%%%%%%%%%%%%%%%%%%%%%%%%%%%%%%%%%%%%%%%%%%%%%%%%%%%%%%%%%%%%%%%%%%%%%%%%%%%%%%%%%%%%%%%%%%%%%%%%%%
\subsection{Paramagnetic phase}
%%%%%%%%%%%%%%%%%%%%%%%%%%%%%%%%%%%%%%%%%%%%%%%%%%%%%%%%%%%%%%%%%%%%%%%%%%%%%%%%%%%%%%%%%%%%%%%%%%%

It is known that breaking both inversion and time-reversal symmetries are required for the appearance of nonreciprocal light propagation.
\cite{Hopfield-1960,Brown-1963,Szaller-2013}
As for directional dichroism, we give a short explanation in Appendix \ref{appendix:directional} for the later convenience.
In the paramagnetic phase above the N\'{e}el temperature,
\Co~has the inversion symmetry even in the presence of an external magnetic field.
Therefore, the directional dichroism does not appear in the paramagnetic phase.
On the other hand, the electric dipole described by the quadrupole operators can cause a unique response to light absorption.
We derive these from a microscopic model for a finite magnetic field parallel to the $z$ direction.

Following Ref. \ref{ref:Matsumoto-2017}, we consider the following local Hamiltonian:
\begin{align}
&\H = \H_0 + \H', \cr
&\H_0 = D (S^z)^2 - g\mu_{\rm B} H_z S^z, \cr
&\H' = - g\mu_{\rm B} \bH^\omega\cdot\bS - \bE^\omega\cdot\bp.
\label{eqn:H-x}
\end{align}
$\H_0$ is the spin Hamiltonian under an external field.
$D(>0)$ represents the easy-plane anisotropy,
$g$ is the $g$-factor, $\mu_{\rm B}$ is the Bohr magneton,
and $H_z$ is the external static magnetic field applied in the $z$ direction.
$\H'$ is the interaction Hamiltonian between the spin and electromagnetic fields ($\bE^\omega$ and $\bH^\omega$)
with $\omega$ as their angular frequency.
$\H'$ leads to optical absorption and we take it as a perturbation.
For \Co, matrix forms of the spin operators for $S=3/2$ are given by Eq. (\ref{eqn:S-mat}).

%%%%%%%%%%%%%%%%%%%%%%%%%%%%%%%%%%%%%%%%%%%%%%%%%%%%%%%%%%%%%%%%%%%%%%%%%%%%%%%%%%%%%%
\begin{figure}[t]
\begin{center}
\includegraphics[width=7cm]{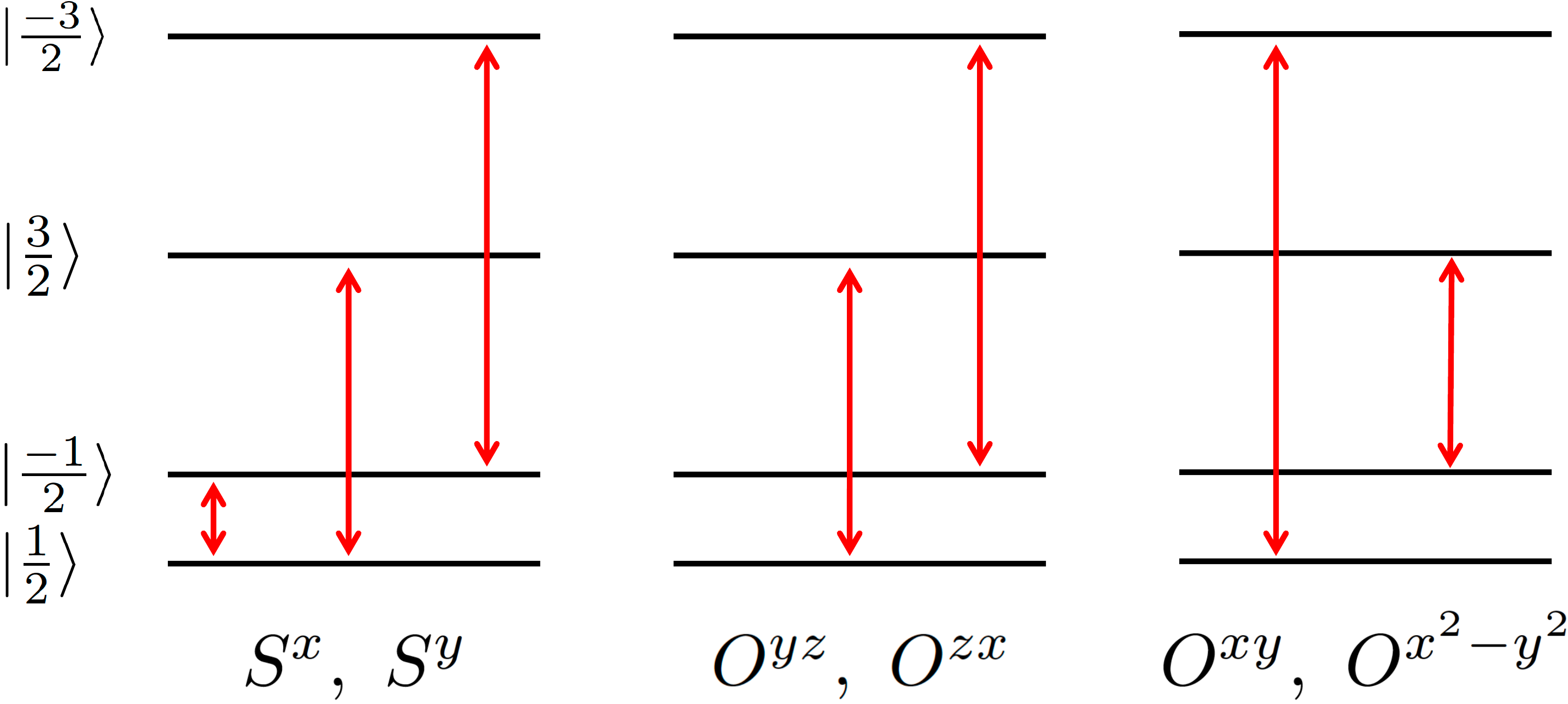}
\end{center}
\caption{
(Color online)
Schematic of energy level scheme for $\bH\parallel z$.
Possible transitions are separately shown for $(S^x,S^y)$, $(O^{yz},O^{zx})$, and $(O^{xy},O^{x^2-y^2})$.
}
\label{fig:level-hz}
\end{figure}
%%%%%%%%%%%%%%%%%%%%%%%%%%%%%%%%%%%%%%%%%%%%%%%%%%%%%%%%%%%%%%%%%%%%%%%%%%%%%%%%%%%%%%

The spin Hamiltonian $\H_0$ in Eq. (\ref{eqn:H-x}) is already diagonal.
The energy eigenstates are expressed as $|m\rangle$ with $m=\pm\frac{1}{2},\pm\frac{3}{2}$ as their $S^z$ values.
For low fields, the energy level scheme is shown in Fig. \ref{fig:level-hz}.
We discuss $\bE^\omega \parallel x$ and $\bH^\omega \parallel y$ configuration.
In this case, $\H'$ is expressed as
\begin{align}
\H'
&= -g\mu_{\rm B} H^\omega S^y - E^\omega p^x
\label{eqn:H'-z} \\
&= -h^\omega S^y - e_1^\omega O^{zx} - e_{1'}^\omega O^{yz} - e_2^\omega O^{x^2-y^2} - e_{2'}^\omega O^{xy},
\nonumber
\end{align}
where
$h^\omega=g\mu_{\rm B}H^\omega$,
$e_1^\omega=K_1 E^\omega$,
$e_{1'}^\omega=K_1' E^\omega$,
$e_2^\omega=K_2 E^\omega$, and
$e_{2'}^\omega=K_2' E^\omega$.
In Eq. (\ref{eqn:H'-z}), we used Eq. (\ref{eqn:p-C3}) for $p^x$.
Matrix forms of the quadrupole operators for $S=3/2$ are given by Eq. (\ref{eqn:O-mat}).
The transition probability from the $m$ to $l$ states is given by the following Fermi's Golden rule:
\begin{align}
W_{m\rightarrow l} = 2\pi \left| \braket{l|\H'|m} \right|^2 \delta(\omega - \Delta E_{lm} ).
\end{align}
Here, $\hbar=1$ and $\Delta E_{lm}=E_l - E_m$ represents the energy difference between the two states.
The energy eigenvalue are given by
$E_{\pm\frac{1}{2}}=\frac{1}{4}D \mp \frac{1}{2} g\mu_{\rm B} H_z$
and
$E_{\pm\frac{3}{2}}=\frac{9}{4}D \mp \frac{3}{2} g\mu_{\rm B} H_z$.
There are two channels in the excitation processes by $\H'$:
magnetic channel via $S^y$ and electric channel via $O^{zx}$, $O^{yz}$, $O^{x^2-y^2}$ and $O^{xy}$, as shown in Eq. (\ref{eqn:H'-z}).
The matrix elements of those operators are given in Eqs. (\ref{eqn:S-mat}) and (\ref{eqn:O-mat}).
The possible excitation processes are summarized in Fig. \ref{fig:level-hz},
and the transition probabilities are given by
\begin{align}
&W_{\frac{1}{2}\rightarrow-\frac{1}{2}} = 2\pi (h^\omega)^2 \delta(\omega - g\mu_{\rm B} H_z), \label{eqn:W-para} \\
&W_{\frac{1}{2}\rightarrow \frac{3}{2}} = 6\pi \left[ \left(\frac{h^\omega}{2}+e_{1'}^\omega\right)^2 + (e_1^\omega)^2 \right]
                                          \delta(\omega - 2D + g\mu_{\rm B} H_z), \cr
&W_{\frac{1}{2}\rightarrow-\frac{3}{2}} = 6\pi \left[ (e_2^\omega)^2 + (e_{2'}^\omega)^2 \right] \delta(\omega - 2D - 2g\mu_{\rm B} H_z).
\nonumber
\end{align}
The $\frac{1}{2}\rightarrow-\frac{1}{2}$ transition is pure magnetic and the resonance frequency is given by $\omega = g\mu_{\rm B} H_z$.
In the $\frac{1}{2}\rightarrow\frac{3}{2}$ transition of $\omega = 2D - g\mu_{\rm B} H_z$,
both the magnetic and electric channels contribute to the transition.
Since the matrix elements of $S^y$ and $O^{yz}$ are pure imaginary,
the $(\frac{h^\omega}{2}+e_{1'}^\omega)^2=(\frac{g\mu_{\rm B}H^\omega}{2}+K_1' E^\omega)^2$ term appears.
The resulting $K_1' H^\omega E^\omega$ term is proportional to $K_1'$
and it has different values when the propagating direction of light is reversed,
i.e. $(E^\omega,H^\omega)\rightarrow(-E^\omega,H^\omega)$ or $(E^\omega,H^\omega)\rightarrow(E^\omega,-H^\omega)$.
This means the directional dichroism.
However, the coefficient $K_1'$ is staggered between the $(a,b')\leftrightarrow(b,a')$ sites,
as in Eqs. (\ref{eqn:p-A}), (\ref{eqn:pB-inverse}), (\ref{eqn:pB-2-fold}), and (\ref{eqn:pA'}).
Owing to the cancellation of these terms, the directional dichroism does not appear in the paramagnetic phase, as expected.

In Eq. (\ref{eqn:W-para}), there is an interesting transition of $\frac{1}{2}\rightarrow-\frac{3}{2}$.
It is pure electric and the resonance frequency is given by $\omega = 2D + 2g\mu_{\rm B} H_z$.
The slope of $H_z$ in $\omega$ is twice as that of the $\frac{1}{2}\rightarrow-\frac{1}{2}$ case, i.e. quadrupolar excitation.
\cite{Akaki-2017}
This is owing to the fact that the $O^{x^2-y^2}$ and $O^{xy}$ quadrupole operators
have a finite matrix element when the $S^z$ values differ $\pm 2$ [see Eq. (\ref{eqn:O-mat})].
Notice that this also results in the $2\theta$ rotation component of the electric dipole, as discussed in the previous section.

%%%%%%%%%%%%%%%%%%%%%%%%%%%%%%%%%%%%%%%%%%%%%%%%%%%%%%%%%%%%%%%%%%%%%%%%%%%%%%%%%%%%%%%%%%%%%%%%%%%
\subsection{Ordered phase}
%%%%%%%%%%%%%%%%%%%%%%%%%%%%%%%%%%%%%%%%%%%%%%%%%%%%%%%%%%%%%%%%%%%%%%%%%%%%%%%%%%%%%%%%%%%%%%%%%%%

In the ordered phase, a spontaneous magnetic moment appears along the $x$ direction.
The moment is staggered between the $(a,a')\leftrightarrow(b,b')$ sites and it breaks the inversion symmetry.
Since the both inversion and time-reversal symmetries are broken,
we can expect various types of dichroism below the N\'{e}el temperature.

%%%%%%%%%%%%%%%%%%%%%%%%%%%%%%%%%%%%%%%%%%%%%%%%%%%%%%%%%%%%%%%%%%%%%%%%%%%%%%%%%%%%%%%%%%%%%%%%%%%
\subsubsection{Symmetry analysis}
%%%%%%%%%%%%%%%%%%%%%%%%%%%%%%%%%%%%%%%%%%%%%%%%%%%%%%%%%%%%%%%%%%%%%%%%%%%%%%%%%%%%%%%%%%%%%%%%%%%

The absorption rate is proportional to the following matrix element:
\begin{align}
|\braket{f| \left( \H_s + \H_p \right) |i}|^2.
\label{eqn:matrix-00}
\end{align}
Here, $|i\rangle$ and $|f\rangle$ are initial and final states, respectively.
We assume that they have no degeneracy.
$\H_s=-g\mu_{\rm B} \bH^\omega\cdot \bS$ and $\H_p= -\bE^\omega\cdot \bP$ are perturbation Hamiltonians
for the magnetic and electric channels, respectively.
Here, $\bS$ and $\bP$ represent total spin and total electric dipole (polarization) operators, respectively.
$\bH^\omega$ and $\bE^\omega$ are magnetic and electric fields of light, respectively.
Since we only consider absorption process in Eq. (\ref{eqn:matrix-00}), $\bH^\omega$ and $\bE^\omega$ are complex for circularly polarized light.

First, we study a case in the absence of the external magnetic field ($\bH=0$).
The AF moment aligns in the easy axis along the $x$ direction.
In this case, there still remains the twofold symmetry around the $y$-axis shown in Fig. \ref{fig:crystal}.
The twofold rotational operation $C_{2y}$ restricts the possible dichroism as follows.
Since the Hamiltonian is invariant under the $C_{2y}$ transformation,
the energy eigenstates are also eigenstates of $C_{2y}$,
i.e. $C_{2y}|i\rangle=\lambda_i|i\rangle$ and $C_{2y}|f\rangle=\lambda_f|f\rangle$ with $|\lambda_i|=|\lambda_f|=1$.
The matrix element in Eq. (\ref{eqn:matrix-00}) is calculated as
\begin{align}
|\braket{f| ( \H_s + \H_p ) |i}|^2
&= |\braket{C_{2y} f| C_{2y} ( \H_s + \H_p ) C_{2y}^{-1} |C_{2y} i}|^2 \cr
&= |\braket{f| C_{2y} ( \H_s + \H_p ) C_{2y}^{-1} |i}|^2.
\label{eqn:matrix-0}
\end{align}
For $C_{2y}$, $\bS$ and $\bP$ are transformed as
$C_{2y}(S^x,S^y,S^z)C_{2y}^{-1}=(-S^x,S^y,-S^z)$ and
$C_{2y}(P^x,P^y,P^z)C_{2y}^{-1}=(-P^x,P^y,-P^z)$, respectively.
\cite{note-P-C2}
This means that the absorption rate is invariant under the following transformation for the electromagnetic fields:
\begin{align}
&(E_x^\omega,E_y^\omega,E_z^\omega)\rightarrow (-E_x^\omega,E_y^\omega,-E_z^\omega), \cr
&(H_x^\omega,H_y^\omega,H_z^\omega)\rightarrow (-H_x^\omega,H_y^\omega,-H_z^\omega).
\label{eqn:trans-1}
\end{align}
This restricts the possible dichroism in the AF($H=0$) phase.
In Tables \ref{table:trans-circular} and \ref{table:trans-linear} in Appendix \ref{appendix:trans},
we list the transformed electromagnetic fields for circularly and linearly polarized lights, respectively.

Next, we discuss a case under a finite external magnetic field along the $z$ direction ($\bH\parallel z$).
In this case, a uniform magnetic moment in the $z$ direction also appears in addition to the AF moment.
Then, the $C_{2y}$ symmetry shown in Fig. \ref{fig:crystal} is broken owing to the $z$ component of the magnetic moment.
Let us consider the symmetry transformation under the field.
The crystal structure of \Co~has a $c$-glide symmetry with respect to the mirror operation perpendicular to the $y$ direction.
\cite{IN-table}
We describe $\sigma_y^c$ as the $c$-glide operation.
We can see that the magnetic structure under the field is invariant under the $\sigma_y^c\Theta$ operation.
Then, the energy eigenstates are also eigenstates of the symmetry transformation of $\sigma_y^c\Theta$.
Notice that the absolute value of the eigenvalue is unity.
In the same way as Eq. (\ref{eqn:matrix-0}), the matrix element is calculated as
\cite{J-J-Sakurai-2017}
\begin{align}
&|\braket{f| ( \H_s + \H_p ) |i}|^2
=|\braket{\Theta i| \Theta ( \H_s^\dagger + \H_p^\dagger ) \Theta^{-1} |\Theta f}|^2 \cr
&=|\braket{\sigma_y^c \Theta i| \sigma_y^c ( -\H_s + \H_p ) (\sigma_y^c)^{-1} |\sigma_y^c \Theta f}|^2 \cr
&=|\braket{i| \sigma_y^c ( -\H_s + \H_p ) (\sigma_y^c)^{-1} |f}|^2 \cr
&=|\braket{f| \sigma_y^c ( -\H_s^\dagger + \H_p^\dagger ) (\sigma_y^c)^{-1} |i}|^2.
\label{eqn:matrix-H}
\end{align}
Here, we used
$\Theta \H_s^\dagger \Theta^{-1}
= \Theta [ -g\mu_{\rm B} (\bH^\omega)^* \cdot\bS ] \Theta^{-1}
= g\mu_{\rm B} \bH^\omega \cdot\bS
=-\H_s$
and
$\Theta \H_p^\dagger \Theta^{-1}
=\Theta [- (\bE^\omega)^*\cdot\bP ] \Theta^{-1}
=- \bE^\omega\cdot\bP
=\H_p$.
For the $c$-glide operation, $\bS$ and $\bP$ are transformed as
$\sigma_y^c(S^x,S^y,S^z)(\sigma_y^c)^{-1}=(S^x,-S^y,S^z)$ and $\sigma_y^c(P^x,P^y,P^z)(\sigma_y^c)^{-1}=(-P^x,P^y,-P^z)$.
\cite{note-P-c-glide}
This means that the absorption rate is invariant under the following transformation:
\begin{align}
&(E_x^\omega,E_y^\omega,E_z^\omega)\rightarrow (-E_x^\omega,E_y^\omega,-E_z^\omega)^*, \cr
&(H_x^\omega,H_y^\omega,H_z^\omega)\rightarrow (-H_x^\omega,H_y^\omega,-H_z^\omega)^*,
\label{eqn:trans-2}
\end{align}
which is equivalent to Eq. (\ref{eqn:trans-1}) except for the complex conjugate operation.
This restricts the possible dichroism in the AF($H=0$) and AF($\bH\parallel z$) phases.
For linearly polarized light, $\bH^\omega$ and $\bE^\omega$ are real and Eqs. (\ref{eqn:trans-1}) and (\ref{eqn:trans-2}) become identical.
In this case, the property of dichroism becomes the same in the AF($H=0$) and AF($\bH\parallel z$) phases
for the linearly polarized light.
It can be different for circularly polarized light.
We list the transformed electromagnetic fields in Tables \ref{table:trans-circular} and \ref{table:trans-linear}.

\begin{figure}[t]
\begin{center}
\includegraphics[width=6.5cm]{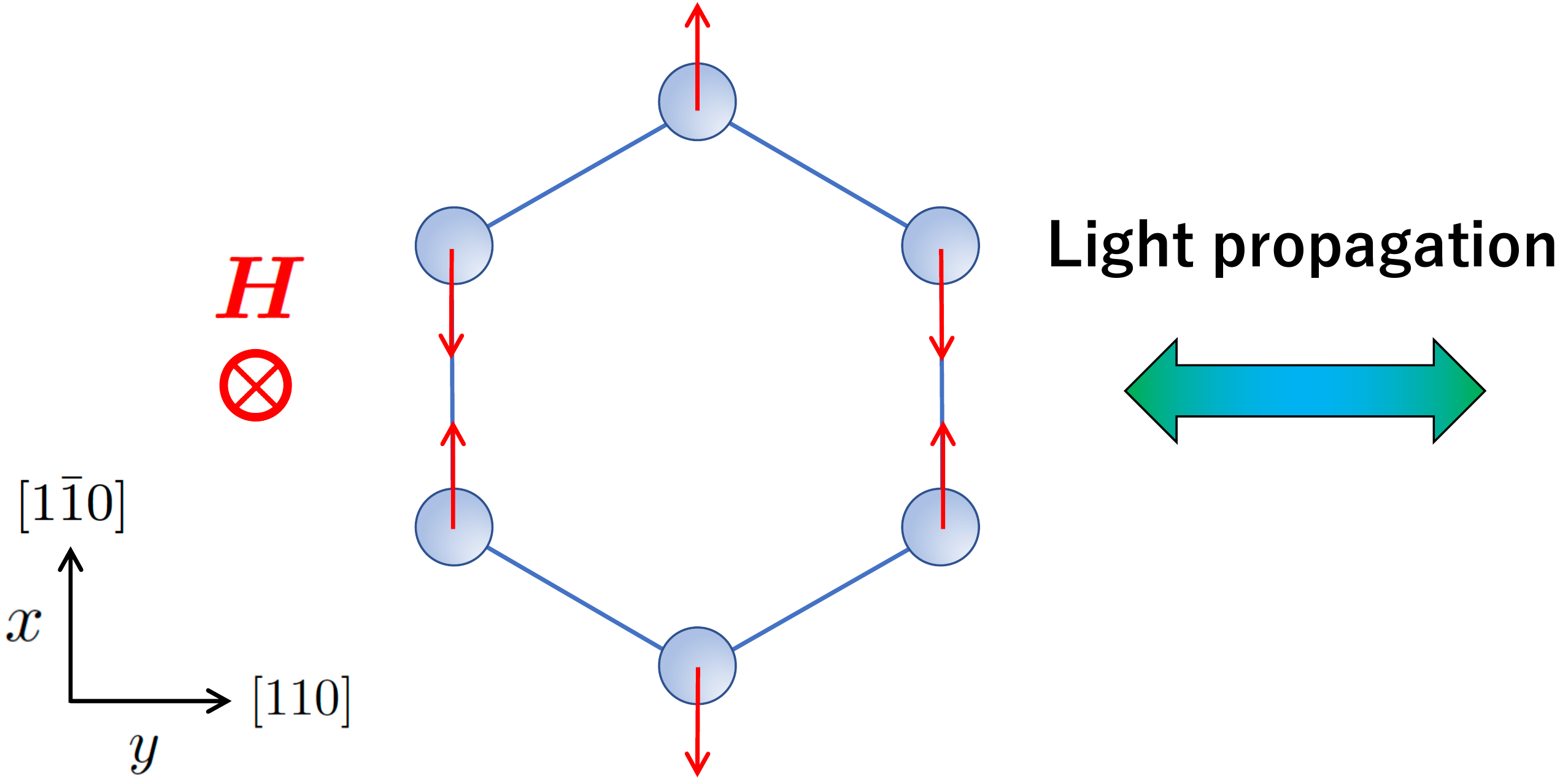}
\end{center}
\caption{
(Color online)
Schematic of light propagation to probe various types of dichroism in the ordered phases of \Co.
We assume that the AF moment aligns in the $x$ direction.
External magnetic field is applied in the $z$ direction.
DD, NCD, and MCD are observable when the light propagates in the $y$ direction,
which is perpendicular to the AF moment and the symmetrical $z$-axis.
}
\label{fig:dichroism}
\end{figure}

From Tables \ref{table:trans-circular} and \ref{table:trans-linear},
optical property of dichroism is obtained and we summarize the result in Table \ref{table:dichroism}.
Strong characteristic property appears when the light propagates in the $y$ direction.
In this case, directional dichroism (DD) always appears for linearly polarized light.
In addition, both natural circular dichroism (NCD) and magnetic circular dichroism (MCD) are observable.
On the other hand, circular dichroism (CD) is unobservable.
In \Co, the $y$ direction is the special one for the optical property,
in which various types of dichroism are observable (see Table \ref{table:dichroism}).
We propose a measurement with a light propagating in the $y$ direction as shown in Fig. \ref{fig:dichroism}.

%%%%%%%%%%%%%%%%%%%%%%%%%%%%%%%%%%%%%%%%%%%%%%%%%%%%%%%%%%%%%%%%%%%%%%%%%%%%%%%%%%%%%%
\begin{table}[t]
\caption{
Observable dichroism in various phases of \Co.
Para, AF$_0$, and AF$_H$ represent the paramagnetic, AF$(H=0)$, and AF$(\bH\parallel z)$ phases, respectively.
Circular dichroism (CD) is defined by $W_{\rm L}(\pm)\neq W_{\rm R}(\pm)$,
where $W_{\rm L}(\pm)$ and $W_{\rm R}(\pm)$ represent the absorption rates for left and right circularly polarized lights
propagating along the $\pm$ directions of a certain axis, respectively.
Natural circular dichroism (NCD) is defined by $W_{\rm L}(\pm)\neq W_{\rm R}(\mp)$.
Magnetic circular dichroism (MCD) is defined by $W_{\rm L}(\pm)\neq W_{\rm L}(\mp)$ and $W_{\rm R}(\pm)\neq W_{\rm R}(\mp)$.
Directional dichroism (DD) is defined by $W(\pm)\neq W(\mp)$,
where $W(\pm)$ represents the absorption rates for linearly polarized lights propagating along the $\pm$ directions of a certain axis, respectively.
$I$, $\Theta$, $C_{2y}$, and $\sigma_y^c$
represent the inversion, time-reversal, twofold rotation around the $y$-axis, and the $c$-glide operations, respectively.
In the Para phase, both $I$ and $\Theta$ remain, which is represented by the symbol ``$\circ$", and the dichroism does not appear.
This is represented by the symbol ``$-$" in the table.
In the ordered phases, in contrast, both $I$ and $\Theta$ are broken,
which are represented by the symbol ``$-$", and various types of dichroism can appear in principal.
The remaining symmetries of $C_{2y}$ and $\sigma_y^c \Theta$, however, restrict the possible dichroism in the ordered phases.
CD is observable in the AF$_H$ phase with a light propagating in the $x$ or $z$ directions,
which are represented by the symbols ``$x$" and ``$z$", respectively.
NCD, MCD, and DD are observable in the ordered phases with a light propagating in the $y$ direction.
This is represented by the symbol ``$y$".
For a light propagating in the $x$ or $z$ directions, MCD and DD are observable as shown below.
In this case, DD becomes exceptionally unobservable when the light is polarized linearly along the principal axes as
$(\bE^\omega,\bH^\omega)=(E_y^\omega \be_y,H_z^\omega \be_z),(E_z^\omega \be_z,H_y^\omega \be_y),
(E_x^\omega \be_x,H_y^\omega \be_y),(E_y^\omega \be_y,H_x^\omega \be_x)$.
Here, $\be_x$ and $\be_y$ represent the unit vectors along the $x$- and $y$-axes, respectively.
In this sense, the symbol $(x,z)$ is used for DD.
In the AF$_0$ phase, $I\Theta$ is also a symmetry transformation, as discussed in Eq. (\ref{eqn:P-AF-0}).
Since this transformation is equivalent to $\sigma_y^c\Theta C_{2y}$ in the matrix element of Eq. (\ref{eqn:matrix-00}),
$I\Theta$ does not alter the result.
}
\begin{tabular}{ccccccccc}
\hline
 & $I$ & $\Theta$ & $C_{2y}$ & $\sigma_y^c\Theta$ & CD & NCD & MCD & DD \cr
\hline 
Para & $\circ$ & $\circ$ & $\circ$ & $\circ$ & $-$ & $-$ & $-$ & $-$ \cr
AF$_0$ & $-$ & $-$ & $\circ$ & $\circ$ & $-$ & $y$ & $y$ & $y,(x,z)$ \cr
AF$_H$ & $-$ & $-$ & $-$ & $\circ$ & $x,z$ & $y$ & $x,y,z$ & $y,(x,z)$ \cr
\hline
\end{tabular}
\label{table:dichroism}
\end{table}
%%%%%%%%%%%%%%%%%%%%%%%%%%%%%%%%%%%%%%%%%%%%%%%%%%%%%%%%%%%%%%%%%%%%%%%%%%%%%%%%%%%%%%

%%%%%%%%%%%%%%%%%%%%%%%%%%%%%%%%%%%%%%%%%%%%%%%%%%%%%%%%%%%%%%%%%%%%%%%%%%%%%%%%%%%%%%
%%%%%%%%%%%%%%%%%%%%%%%%%%%%%%%%%%%%%%%%%%%%%%%%%%%%%%%%%%%%%%%%%%%%%%%%%%%%%%%%%%%%%%

%%%%%%%%%%%%%%%%%%%%%%%%%%%%%%%%%%%%%%%%%%%%%%%%%%%%%%%%%%%%%%%%%%%%%%%%%%%%%%%%%%%%%%%%%%%%%%%%%%%
\subsubsection{Analysis of directional dichroism with microscopic model}
\label{sec:micro}
%%%%%%%%%%%%%%%%%%%%%%%%%%%%%%%%%%%%%%%%%%%%%%%%%%%%%%%%%%%%%%%%%%%%%%%%%%%%%%%%%%%%%%%%%%%%%%%%%%%

%%%%%%%%%%%%%%%%%%%%%%%%%%%%%%%%%%%%%%%%%%%%%%%%%%%%%%%%%%%%%%%%%%%%%%%%%%%%%%%%%%%%%%
\begin{figure}[t]
\begin{center}
\includegraphics[width=4.7cm]{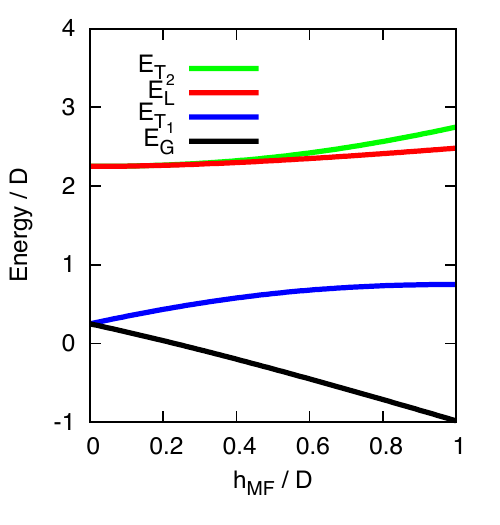}
\end{center}
\caption{
(Color online)
Molecular field $h_{\rm MF}$ dependence of the energy eigenstates.
Here, the molecular field is along the $x$ direction.
$E_\rG$ is the ground state energy, and $E_m$ ($m=\rT_1,\rL,\rT_2$) are the energies for the excited states.
}
\label{fig:energy-hx}
\end{figure}
%%%%%%%%%%%%%%%%%%%%%%%%%%%%%%%%%%%%%%%%%%%%%%%%%%%%%%%%%%%%%%%%%%%%%%%%%%%%%%%%%%%%%%

%%%%%%%%%%%%%%%%%%%%%%%%%%%%%%%%%%%%%%%%%%%%%%%%%%%%%%%%%%%%%%%%%%%%%%%%%%%%%%%%%%%%%%
\begin{figure}[t]
\begin{center}
\includegraphics[width=6cm]{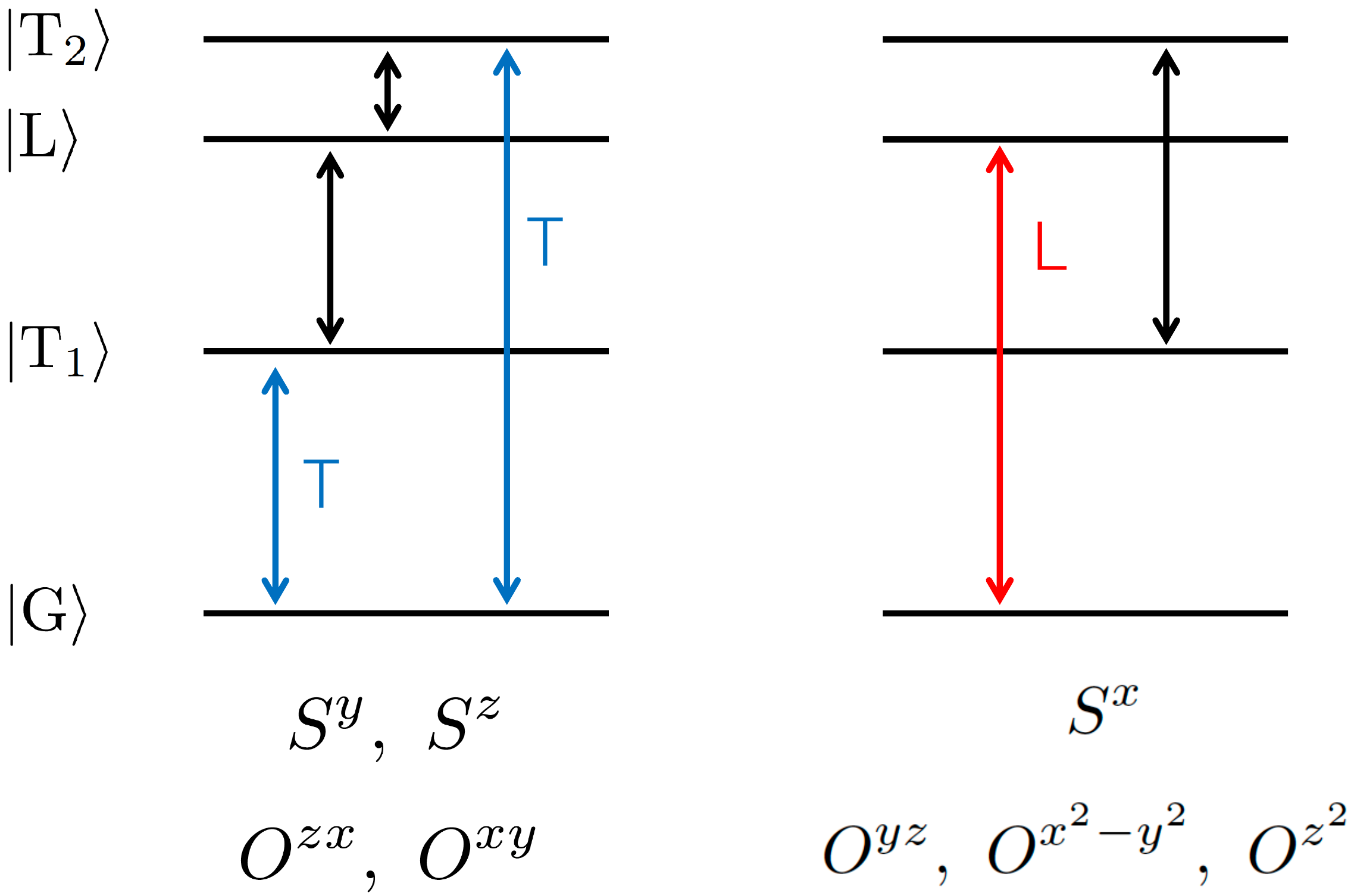}
\end{center}
\caption{
(Color online)
Schematic of energy level scheme under the molecular field in the $x$ direction.
Nonzero transitions by the operators are shown for transverse ($S^y$, $S^z$, $O^{zx}$, $O^{xy}$)
and longitudinal ($S^x$, $O^{yz}$, $O^{x^2-y^2}$, $O^{z^2}$) excitations.
A magnetic moment is induced in the $x$ direction under the molecular field.
For the transverse excitations, the $|\rT_1\rangle$ and $|\rT_2\rangle$ states are excited from the ground state
by the transverse spin components ($S^y$ and $S^z$).
On the other hand, for the longitudinal excitation,
the $|\rL\rangle$ state is excited by the longitudinal spin component ($S^x$), as in the \Ba~case.
\cite{Penc-2012,Romhanyi-2012,Soda-2014,Soda-2018}
}
\label{fig:level-hx}
\end{figure}
%%%%%%%%%%%%%%%%%%%%%%%%%%%%%%%%%%%%%%%%%%%%%%%%%%%%%%%%%%%%%%%%%%%%%%%%%%%%%%%%%%%%%%

In this sub-subsection, we focus on the light propagating in the special $y$ direction specific to \Co.
We discuss the directional dichroism within a microscopic model and consider the following local Hamiltonian for $\bH=0$:
\begin{align}
\H_0 = D(S^z)^2 - h_{\rm MF} S^x.
\end{align}
Here, $h_{\rm MF}$ represents the molecular field.
$\H_0$ can be diagonalized analytically for $S=3/2$.
The $h_{\rm MF}$ dependence of the energy eigenstates are shown in Fig. \ref{fig:energy-hx}.
They are expressed as $|m\rangle$ ($m=\rG,\rT_1,\rL,\rT_2$),
where L and T represent the longitudinal and transverse excited states, respectively.
There is a longitudinal excited state, as in \Ba.
\cite{Penc-2012,Romhanyi-2012,Soda-2014,Soda-2018}
The energy level scheme is shown in Fig. \ref{fig:level-hx}.
We notice that the local Hamiltonian $\H_0$ is invariant under the $\pi$ rotation around the $x$-axis, namely $C_{2x}$.
Since the spin and quadrupole operators are transformed as
$C_{2x}(S^x,S^y,S^z)C_{2x}^{-1}=(S^x,-S^y,-S^z)$
and 
$C_{2x}(O^{yz},O^{zx},O^{xy},O^{x^2-y^2},O^{z^2})C_{2x}^{-1}=(O^{yz},-O^{zx},-O^{xy},O^{x^2-y^2},O^{z^2})$,
they are classified in two different groups as
$(S^y,S^z,O^{zx},O^{xy})$ and $(S^x,O^{yz},O^{x^2-y^2},O^{z^2})$,
\cite{note-classification}
as shown in Fig. \ref{fig:level-hx}.

First, we study the $(E_z^\omega,H_x^\omega)$ configuration.
This is expressed by the following perturbation Hamiltonian [see Eq. (\ref{eqn:p-C3})]:
\begin{align}
\H' = -h^\omega S^x - e_3^\omega O^{z^2},
\end{align}
Here, $h^\omega=g\mu_{\rm B}H_x^\omega$ and $e_3=K_3 E_z^\omega$.
As shown in Fig. \ref{fig:level-hx}, the longitudinal state $|\rL\rangle$ is excited
by the both $S^x$ and $O^{z^2}$ operators in $\H'$.
The transition probability is calculated as
\begin{align}
&W_{\rG\rightarrow\rL}
= 2\pi \left| \braket{\rL|\H'|\rG} \right|^2 \delta(\omega - \Delta E_{\rL\rG} ) \cr
&= 2\pi
\Bigl[
  f_{mm}(h_{\rm MF})(h^\omega)^2 + f_{ee}(h_{\rm MF}) (e_3^\omega)^2 \cr
&~~~~~~~~~
+ f_{me}(h_{\rm MF}) h^\omega e_3^\omega
\Bigr] \delta(\omega - \Delta E_{\rL\rG}),
\label{eqn:W-hx}
\end{align}
with
\begin{align}
&f_{mm}(h_{\rm MF}) = \left| \braket{\rL|S^x|\rG} \right|^2, \cr
&f_{ee}(h_{\rm MF}) = \left| \braket{\rL|O^{z^2}|\rG} \right|^2, \cr
&f_{me}(h_{\rm MF}) = \braket{\rG|S^x|\rL} \braket{\rL|O^{z^2}|\rG} + {\rm h.c.}
\label{eqn:f-me}
\end{align}
Here, $\Delta E_{\rL\rG}=E_\rL - E_\rG=2\sqrt{D^2 + Dh_{\rm MF} + h_{\rm MF}^2}$ is the energy difference
between the $|\rL\rangle$ and $|\rG\rangle$ states.
$f_{mm}(h_{\rm MF})$ and $f_{ee}(h_{\rm MF})$ represent the intensity for the pure magnetic and electric channels, respectively.
In Eq. (\ref{eqn:f-me}), the $f_{me}(h_{\rm MF})$ term represents an interference between the magnetic and electric channels.
The characteristic point is that $f_{me}(h_{\rm MF})$ is an odd function of $h_{\rm MF}$,
i.e. $f_{me}(-h_{\rm MF})=-f_{me}(h_{\rm MF})$.
This comes from the fact that the matrix element of $\braket{\rG|S^x|\rL}$ is an odd function of $h_{\rm MF}$,
whereas $\braket{\rL|O^{z^2}|\rG}$ is an even function.
\cite{note-sign-hx}

%%%%%%%%%%%%%%%%%%%%%%%%%%%%%%%%%%%%%%%%%%%%%%%%%%%%%%%%%%%%%%%%%%%%%%%%%%%%%%%%%%%%%%
\begin{table}[t]
\caption{
Relation of coupling constant $K_n$ $(n=1,2,1',2',3)$ and matrix elements of operators between various ($a,b,a',b'$) sites.
We chose the value at the $a$ site as the standard value.
The relations for $K_n$ is taken from Eqs. (\ref{eqn:p-A}), (\ref{eqn:pB-inverse}), (\ref{eqn:pB-2-fold}), and (\ref{eqn:pA'}).
The matrix elements mean $\braket{\rG|S|m}$ and $\braket{m|O|\rG}$ for the spin and quadrupole operators, respectively.
Here, $m={\rm L}$ for the L-mode, whereas $m={\rm T}_1$ or ${\rm f}={\rm T}_2$ for the T-mode.
The symmetry analysis of the matrix elements is given in Ref. \ref{ref:note-sign-hx}.
After the summation over the four sites, the $f_{me}(h_{\rm MF})$ term in Eq. (\ref{eqn:W-hx}) remains
in the $(K_3 O^{z^2},S^x)$ combination for the L-mode.
This is consistent with the $(E_z^\omega,H_x^\omega)$ configuration.
For the T-mode, it remains in the $(K_1 O^{zx},S^z)$ combination and is consistent with the $(E_x^\omega,H_z^\omega)$ configuration.
The $f_{me}(h_{\rm MF})$ term vanishes for other combinations of $K_n O$ and $S$, owing to the cancellation over the four sites.
}
\begin{tabular}{ccccc}
\hline
 & $a$ & $b$ & $a'$ & $b'$ \cr
\hline
$K_1$ & $+$ & $-$ & $+$ & $-$ \cr
$K_2$ & $+$ & $-$ & $+$ & $-$ \cr
$K_1'$ & $+$ & $-$ & $-$ & $+$ \cr
$K_2'$ & $+$ & $-$ & $-$ & $+$ \cr
$K_3$ & $+$ & $-$ & $+$ & $-$ \cr
\hline
$O^{zx}$ & $+$ & $-$ & $+$ & $-$ \cr
$O^{x^2-y^2}$ & $+$ & $+$ & $+$ & $+$ \cr
$O^{yz}$ & $+$ & $-$ & $+$ & $-$ \cr
$O^{xy}$ & $+$ & $+$ & $+$ & $+$ \cr
$O^{z^2}$ & $+$ & $+$ & $+$ & $+$ \cr
\hline
$K_1 O^{zx}$ ($p^x$, T-mode) & $+$ & $+$ & $+$ & $+$ \cr
$K_2 O^{x^2-y^2}$ ($p^x$, L-mode) & $+$ & $-$ & $+$ & $-$ \cr
$K_1' O^{yz}$ ($p^x$, L-mode) & $+$ & $+$ & $-$ & $-$ \cr
$K_2' O^{xy}$ ($p^x$, T-mode) & $+$ & $-$ & $-$ & $+$ \cr
\hline
$K_3 O^{z^2}$ ($p^z$, L-mode) & $+$ & $-$ & $+$ & $-$ \cr
\hline
$S^x$ (L-mode) & $+$ & $-$ & $+$ & $-$ \cr
$S^y$ (T-mode) & $+$ & $-$ & $+$ & $-$ \cr
$S^z$ (T-mode) & $+$ & $+$ & $+$ & $+$ \cr
\hline
\end{tabular}
\label{table:matrix-T-mode}
\end{table}
%%%%%%%%%%%%%%%%%%%%%%%%%%%%%%%%%%%%%%%%%%%%%%%%%%%%%%%%%%%%%%%%%%%%%%%%%%%%%%%%%%%%%%

In the ordered phase, the molecular fields are staggered between the A$(a,a')$ and B$(b,b')$ sites.
This means that $h_{\rm MF}$ in Eq. (\ref{eqn:W-hx}) is reversed as $h_{\rm MF}\rightarrow -h_{\rm MF}$ on the B site.
Since $f_{me}(h_{\rm MF})$ is an odd function of $h_{\rm MF}$ and $e_3^\omega$ ($K_3$) is staggered between the two sites,
the interference term does not cancel in the ordered phase, as expected by the symmetry analysis.
This point is summarized in Table \ref{table:matrix-T-mode} for the L-mode.
After adding the contributions from the four sites, we obtain
\begin{align}
&W_{\rG\rightarrow\rL}(a)+W_{\rG\rightarrow\rL}(a') + W_{\rG\rightarrow\rL}(b)+W_{\rG\rightarrow\rL}(b') \cr
&= 8\pi
\Bigl[
  f_{mm}(h_{\rm MF})(h^\omega)^2 + f_{ee}(h_{\rm MF})(e_3^\omega)^2 \cr
&~~~~~~~~~
  + f_{me}(h_{\rm MF}) h^\omega e_3^\omega
\Bigr] \delta(\omega - \Delta E_{\rL\rG}).
\label{eqn:W-order}
\end{align}
Here, $f_{me}(h_{\rm MF}) h^\omega e_3^\omega$ is the value at the $a$ site.
This term represents the directional dichroism.
In the present discussion, we focused on the Co(1) site.
The same argument holds for the Co(2) site with different coupling constants $K_n$ $(n=1,2,1',2',3)$.

The above discussion is based on the isolated picture of a Co$^{2+}$ ion.
In the ordered phase, spin-wave excitations are formed by the intersite interactions.
\cite{Deng-2018}
For more precise analysis, we have to take this point into account.
We consider then the following perturbation Hamiltonian:
\begin{align}
\H' = - \sum_i \sum_\mu \left( h^\omega S_{i\mu}^x + e_{\mu 3}^\omega O_{i\mu}^{z^2} \right).
\label{eqn:H'-total}
\end{align}
Here, $i$ represents the $i$th unit cell and $\mu$ represents
the $\mu(=a_1,a_1',a_2,a_2',b_1,b_1',b_2,b_2')$ cite in the unit cell (see Fig. \ref{fig:crystal}).
The important point is that both the matrix element $\braket{\rL|S_{i\mu}^x|\rG}$
and the coupling constant $e_{\mu 3}^\omega=K_{\mu 3}E^\omega$ are staggered
on the A $(\mu=a_1,a_1',a_2,a_2')$ and B $(\mu=b_1,b_1',b_2,b_2')$ sites.
There are eight longitudinal spin-wave excitation modes in the first Brillouin zone, reflecting the eight Co sites in the unit cell.
Each L-mode has a finite excitation gap.
The perturbation Hamiltonian in Eq. (\ref{eqn:H'-total}) connects the ground sate to a L-mode
at the $\Gamma$ point ($\bq=0$) in the first Brillouin zone.
Since the matrix element $\braket{\rL| ( h^\omega S_{i\mu}^x +e_{\mu 3}^\omega O_{i\mu}^{z^2} ) |\rG}$
is staggered on the A and B sites, the excited L-mode must be an optical one.
As discussed in the isolated spin picture, this excitation causes the directional dichroism.
In the paramagnetic phase above the N\'{e}el temperature, the dichroism disappears as the AF moment vanishes.

Next, we study the $(E_x^\omega,H_z^\omega)$ configuration.
As shown in Fig. \ref{fig:level-hx}, this configuration probes the T-mode.
The perturbation Hamiltonian is given by
\begin{align}
\H' = -h^\omega S^z - e_1^\omega O^{zx} - e_2^\omega O^{x^2-y^2} - e_{1'}^\omega O^{yz} - e_{2'}^\omega O^{xy}.
\end{align}
Here, $h^\omega=g\mu_{\rm B}H_z^\omega$ and $e_n^\omega=K_n E_x^\omega$ ($n=1,2,1',2'$).
We can discuss the directional dichroism by using Eqs. (\ref{eqn:W-hx}) and (\ref{eqn:f-me})
with the replacement of $\rL\rightarrow\rT_m$ ($m=1,2$).
Relation of matrix elements of the spin and quadrupole operators
between various spin sites are summarized in Table \ref{table:matrix-T-mode}.
For the $(E_x^\omega,H_z^\omega)$ configuration, the matrix element $\braket{\rG|S^z|\rT_m}$ is an even function of $h_{\rm MF}$,
whereas $\braket{\rT_m|O^{zx}|\rG}$ is an odd function of $h_{\rm MF}$.
\cite{note-sign-hx}
After the summation over the four sites, we can check from Table \ref{table:matrix-T-mode}
that the $f_{me}(h_{\rm MF})$ term in Eq. (\ref{eqn:W-order}) remains in the ($K_1 O^{zx},S^z$) combination for the T-mode.
Then, we obtain the following term for the dichroism in the absorption rate:
\begin{align}
8\pi f_{me}(h_{\rm MF}) h^\omega e_1^\omega \delta(\omega - \Delta E_{\rT_m\rG}).
\label{eqn:W-order-T-mode}
\end{align}
Here, $f_{me}(h_{\rm MF})= \braket{\rG|S^z|\rT_m} \braket{\rT_m|O^{zx}|\rG} + {\rm h.c.}$
and $h^\omega e_1^\omega$ represent the values at the $a$ site.
$\Delta E_{\rT_m\rG}$ $(m=1,2$) is the excitation energy given by
\begin{align}
\Delta E_{\rT_m\rG}=h_{\rm MF}&+\sqrt{D^2+Dh_{\rm MF}+h_{\rm MF}^2} \cr
&\mp \sqrt{D^2-Dh_{\rm MF}+h_{\rm MF}^2}.
\end{align}
Here, the $\mp$ are for $m=1$ and $2$, respectively.

Under a finite magnetic field for $\bH\parallel z$, the L- and T-modes are mixed.
The energy eigenvalues of the L- and T-modes at $\bH=0$
continuously evolve with the applied magnetic field from the values at $\bH=0$.
The former and the latter modes are detectable by electron spin resonance measurements
with the $(E_z^\omega,H_x^\omega)$ and $(E_x^\omega,H_z^\omega)$ configurations under $\bH\parallel z$, respectively.

Under a strong field, the magnetic moment saturates and the $|\frac{3}{2}\rangle$ state
becomes the local ground state against the $|\frac{1}{2}\rangle$ state.
In this case, we can also expect pure electric (quadrupolar) excitation by the light absorption
in the $(E_x^\omega,H_y^\omega)$ configuration, as in the paramagnetic phase.
As shown in Fig. \ref{fig:level-hz}, the $\frac{3}{2}\rightarrow -\frac{1}{2}$ transition
is caused by the $O^{x^2-y^2}$ and $O^{xy}$ quadrupoles.
According to the isolated spin picture as in Eq. (\ref{eqn:W-para}), the transition probability is given by
\begin{align}
W_{\frac{3}{2}\rightarrow-\frac{1}{2}} = 6\pi \left[ (e_2^\omega)^2 + (e_{2'}^\omega)^2 \right] \delta(\omega + 2D - 2g\mu_{\rm B} H_z).
\end{align}
The slope of $H_z$ in the resonance frequency $\omega$ of the quadrupolar excitation
is expected to be twice as that of the $\frac{3}{2}\rightarrow \frac{1}{2}$ transition,
as observed in the \Sr~case.
\cite{Akaki-2017}

%%%%%%%%%%%%%%%%%%%%%%%%%%%%%%%%%%%%%%%%%%%%%%%%%%%%%%%%%%%%%%%%%%%%%%%%%%%%%%%%%%%%%%%%%%%%%%%%%%%
\section{Summary and Discussions}
%%%%%%%%%%%%%%%%%%%%%%%%%%%%%%%%%%%%%%%%%%%%%%%%%%%%%%%%%%%%%%%%%%%%%%%%%%%%%%%%%%%%%%%%%%%%%%%%%%%

We investigated the spin-dependent electric polarization appearing in \Co~on the basis of
the symmetry analyses of the Co$^{2+}$ ions in the \Pc~space group by the following procedures.
(i) Each Co$^{2+}$ ion carries spin-dependent electric dipole operator.
The spin dependence is expressed with quadrupole operators.
It is classified by the local point-group symmetry with coupling constants for the quadrupoles
and shown in Eq. (\ref{eqn:p-C3}) for the $C_3$ point-group symmetry.
\cite{Matsumoto-2017}
(ii) For \Pc, there are inversion centers and twofold axes in the unit cell.
The spin dependence of the electric dipole operator at each equivalent Co site is determined
by the transformations with respect to the inversion center and the twofold axis with the common coupling constants.
(iii) The electric polarization is obtained by adding the all contributions from the expectation values of the electric dipole operators at the Co sites.

For \Co, the electric polarization rotates in the basal $ab$-plane with the rotation of the external magnetic field applied in the $ab$-plane.
There are two components.
One rotates in the same direction at the same speed of the rotation of the external field ($\theta$-rotation),
whereas the other rotates in the opposite direction at the twice speed ($2\theta$-rotation).
It is the universal property to have the two rotation components in the presence of the threefold rotational axis,
since the both components rotate by $120^\circ$ after the $120^\circ$ rotation of the external field.
The $2\theta$-rotation component well explains the experimental result.
The $\theta$-rotation component appears when the external magnetic field inclines toward the $c$-axis from the $ab$-plane,
which can be explored by future measurements.

As demonstrated in \Co, the symmetry analysis of the spin-dependent electric dipole operator
is very powerful for understanding magnetoelectric effects in specific materials.
We emphasize that the procedures (i)-(iii) mentioned above
are useful for various quantum spin systems showing magnetoelectric effects
when the magnetic ion with $S\ge 1$ possessing the quadrupole degrees of freedom
occupies a site lacking the inversion symmetry.
The advantage of the theory is that the essential spin-dependence of the electric dipole operator
can be grasped by the symmetry analysis of the space group (including the point group at each magnetic-ion site)
without going into its microscopic origin.
It is useful not only for the electric polarization but also to know other related magnetoelectric effects such as optical properties.
In \Co, we suggest that it is worthwhile to confirm the quadrupolar excitations
and various types of observable dichroism summarized in Table \ref{table:dichroism} in future experiments.

%%%%%%%%%%%%%%%%%%%%%%%%%%%%%%%%%%%%%%%%%%%%%%%%%%%%%%%%%%%%%%%%%%%%%%%%%%%%%%%%%%%%%%%%%%%%%%%%%%%
%\acknowledgments
{\footnotesize\paragraph{\footnotesize Acknowledgments}
We would like to express our sincere thanks to S. Kimura, H. Kusunose, and Y. Yanagi
for stimulating discussions on magnetoelectric effects.
On optical properties related to dichroism, we would like to thank S. Kimura for fruitful discussions.
This work was supported by JSPS KAKENHI Grant Number 17K05516.
}
%%%%%%%%%%%%%%%%%%%%%%%%%%%%%%%%%%%%%%%%%%%%%%%%%%%%%%%%%%%%%%%%%%%%%%%%%%%%%%%%%%%%%%%%%%%%%%%%%%%

%%%%%%%%%%%%%%%%%%%%%%%%%%%%%%%%%%%%%%%%%%%%%%%%%%%%%%%%%%%%%%%%%%%%%%%%%%%%%%%%%%%%%%%%%%%%%%%%%%%
\appendix
\setcounter{equation}{0}
%%%%%%%%%%%%%%%%%%%%%%%%%%%%%%%%%%%%%%%%%%%%%%%%%%%%%%%%%%%%%%%%%%%%%%%%%%%%%%%%%%%%%%%%%%%%%%%%%%%

%%%%%%%%%%%%%%%%%%%%%%%%%%%%%%%%%%%%%%%%%%%%%%%%%%%%%%%%%%%%%%%%%%%%%%%%%%%%%%%%%%%%%%%%%%%%%%%%%%%
\section{Expectation Values of Spin and Quadrupole Operators under Magnetic Field}
\label{appendix:general-S}
%%%%%%%%%%%%%%%%%%%%%%%%%%%%%%%%%%%%%%%%%%%%%%%%%%%%%%%%%%%%%%%%%%%%%%%%%%%%%%%%%%%%%%%%%%%%%%%%%%%

Let us begin with the following local Hamiltonian for a spin under a magnetic field $(H_x,0,H_z)$:
\begin{align}
\H_0 = D (S^z)^2 - g\mu_{\rm B} ( H_x S^x + H_z S^z ).
\label{eqn:H-local}
\end{align}
The magnetic field is applied in the $zx$-plane.
In the absence of the field, the system has a rotational symmetry around the $z$-axis.
Under the finite field, the ground state has no degeneracy.
With respect to the time reversal operation $\Theta$,
spin operators are transformed as $\Theta (S^x,S^y,S^z) \Theta^{-1}=(-S^x,-S^y,-S^z)$.
We also consider a mirror operation perpendicular to the $y$-axis.
It is a unitary transformation and is denoted by $\sigma_y$.
For the mirror operation, the spin operators are transformed as $\sigma_y (S^x,S^y,S^z) \sigma_y^{-1}=(-S^x,S^y,-S^z)$.
We can see that the Hamiltonian is invariant under the $\sigma_y \Theta$ transformation,
i.e. $(\sigma_y\Theta) \H_0 (\sigma_y\Theta)^{-1} = \H_0$.
The ground state $|g\rangle$ is then an eigen state of $\sigma_y \Theta$ as
\begin{align}
\sigma_y \Theta |g\rangle = \lambda |g\rangle.
\label{eqn:lambda}
\end{align}
Here, $\lambda$ is the eigenvalue.
Since $\sigma_y\Theta$ is not Hermitian, $\lambda$ can be complex.

We next calculate the expectation value of $S^x$ as follows:
\cite{J-J-Sakurai-2017}
\begin{align}
\braket{g|S^x|g}
&= \braket{\Theta g|\Theta S^x \Theta^{-1} |\Theta g} \cr
&= -\braket{\Theta g| S^x |\Theta g} \cr
&= -\braket{\sigma_y \Theta g| \sigma_y S^x \sigma_y^{-1} |\sigma_y \Theta g} \cr
&= |\lambda|^2 \braket{g| S^x |g}.
\label{eqn:Sx-0}
\end{align}
This assures $|\lambda|^2=1$.
Equation (\ref{eqn:Sx-0}) also holds when we replace $S^x$ with $S^z$.
As for $S^y$, we obtain
\begin{align}
\braket{g|S^y|g}
= - \braket{g| S^y |g},
\label{eqn:Sy}
\end{align}
owing to $\sigma_y S^y \sigma_y^{-1}=S^y$.
This indicates that $\braket{g|S^y|g}= 0$.
On the other hand, the expectation values of $S^x$ and $S^z$ can be finite.
Therefore, the spin lies is in the $zx$-plane for $|g\rangle$.
This is a natural result for the ground state of the Hamiltonian Eq. (\ref{eqn:H-local}).

With respect to the time reversal operation, quadrupole operators are transformed as $\Theta O^m \Theta^{-1}=O^m$.
As for the mirror operation, they are transformed as
$\sigma_y (O^{yz},O^{zx},O^{xy},O^{x^2-y^2},O^{z^2})\sigma_y^{-1}=(-O^{yz},O^{zx},-O^{xy},O^{x^2-y^2},O^{z^2})$.
In the same manner as Eqs. (\ref{eqn:Sx-0}) and (\ref{eqn:Sy}), we obtain $\braket{g|O^{yz}|g}=\braket{g|O^{xy}|g}=0$.
Notice that the expectation values of the other components of the quadrupole operators can be finite.

We next consider spin rotation around the $z$-axis.
When the spin rotates around the $z$-axis by angle $\phi_s$ from the $|g\rangle$ state, the corresponding state is given by
\begin{align}
|\phi_s\rangle = e^{-i S^z \phi_s} |g\rangle
\end{align}
in $\hbar=1$ unit.
With respect to the unitary transformation $U=e^{-i S^z \phi_s}$, the operators are transformed as
\begin{align}
&U^\dagger
\begin{pmatrix}
S^x \cr
S^y \cr
S^z
\end{pmatrix} 
U
=
\begin{pmatrix}
\cos\phi_s S^x - \sin\phi_s S^y \cr
\cos\phi_s S^y + \sin\phi_s S^x \cr
S^z
\end{pmatrix}, \\
&U^\dagger
\begin{pmatrix}
O^{yz} \cr
O^{zx} \cr
O^{xy} \cr
O^{x^2-y^2} \cr
O^{z^2}
\end{pmatrix} 
U
=
\begin{pmatrix}
\cos\phi_s O^{yz} + \sin\phi_s O^{zx} \cr
\cos\phi_s O^{zx} - \sin\phi_s O^{yz} \cr
\cos{2\phi_s} O^{xy} + \sin{2\phi_s} O^{x^2-y^2} \cr
\cos{2\phi_s} O^{x^2-y^2} - \sin{2\phi_s} O^{xy} \cr
O^{z^2}
\end{pmatrix}. \nonumber
\end{align}
After the spin rotation, expectation value of an operator $A$ is calculated as
$\braket{\phi_s | A | \phi_s} = \braket{Ug| A | Ug} = \braket{g| U^\dagger A U | g}$.
In case of the spin and quadrupole operators, we obtain
\begin{align}
&
\begin{pmatrix}
\braket{S^x} \cr
\braket{S^y} \cr
\braket{S^z}
\end{pmatrix} 
=
\begin{pmatrix}
\cos\phi_s \braket{S^x}_g \cr
\sin\phi_s \braket{S^y}_g \cr
\braket{S^z}_g
\end{pmatrix}, \cr
&
\begin{pmatrix}
\braket{O^{yz}} \cr
\braket{O^{zx}} \cr
\braket{O^{xy}} \cr
\braket{O^{x^2-y^2}} \cr
\braket{O^{z^2}}
\end{pmatrix}
=
\begin{pmatrix}
\sin\phi_s \braket{O^{zx}}_g \cr
\cos\phi_s \braket{O^{zx}}_g \cr
\sin{2\phi_s} \braket{O^{x^2-y^2}}_g \cr
\cos{2\phi_s} \braket{O^{x^2-y^2}}_g \cr
\braket{O^{z^2}}_g
\end{pmatrix}.
\label{eqn:rotation}
\end{align}
Here, the expectation values represent $\braket{\cdots}=\braket{\phi_s|\cdots|\phi_s}$ and $\braket{\cdots}_g=\braket{g|\cdots|g}$,
and we used $\braket{S^y}_g=\braket{O^{yz}}_g=\braket{O^{xy}}_g=0$.
When the spin is rotated, Eq. (\ref{eqn:rotation}) indicates that expectation values become
$\braket{S^x}\propto\cos\phi_s$,
$\braket{S^y}\propto\sin\phi_s$,
$\braket{S^z}=$constant,
$\braket{O^{zx}}\propto\cos\phi_s$,
$\braket{O^{yz}}\propto\sin\phi_s$,
$\braket{O^{x^2-y^2}}\propto\cos{2\phi_s}$,
$\braket{O^{xy}}\propto\sin{2\phi_s}$, and
$\braket{O^{z^2}}=$constant.
These relations on the spin operators hold for general values of spin $S$.
On the other hand, the relations on the quadrupole operators hold for $S\ge 1$ spin systems having the quadrupole degrees of freedom.
For $S=1/2$, notice that the quadrupole operators vanish.

%%%%%%%%%%%%%%%%%%%%%%%%%%%%%%%%%%%%%%%%%%%%%%%%%%%%%%%%%%%%%%%%%%%%%%%%%%%%%%%%%%%%%%%%%%%%%%%%%%%
\section{Spin-Dependent Electric Dipole for Cubic Point Group}
\label{appendix:cubic}
%%%%%%%%%%%%%%%%%%%%%%%%%%%%%%%%%%%%%%%%%%%%%%%%%%%%%%%%%%%%%%%%%%%%%%%%%%%%%%%%%%%%%%%%%%%%%%%%%%%

Spin-dependent electric dipole can be present in the absence of the inversion symmetry.
In cubic systems, $T$ and $T_d$ point group symmetries match this condition and the electric dipole is given by
\cite{Matsumoto-2017}
\begin{align}
p^x = K O^{yz},~~~p^y = K O^{zx},~~~p^z = K O^{xy},
\label{eqn:p-cubic-0}
\end{align}
where $(x,y,z) = x \be_x + y \be_y + z \be_z$ and $\be_i$ ($i = x,y,z$) are the unit vectors in the principal axes:
$\be_x = (1,0,0)$, $\be_y = (0,1,0)$, and $\be_z = (0,0,1)$.
The common constant $K$ represents the equivalency of the $x$-, $y$-, and $z$-axes in the cubic symmetry.
For the later convenience, we rewrite Eq. (\ref{eqn:p-cubic-0}) in the following form:
$p^\alpha = S^\beta K^\alpha_{\beta\gamma} S^\gamma$.
Here, $K^\alpha_{\beta\gamma}$ are third-rank tensors given by
\begin{align}
&K^x =
\begin{pmatrix}
0 & 0 & 0 \cr
0 & 0 & K \cr
0 & K & 0
\end{pmatrix}, \cr
&K^y =
\begin{pmatrix}
0 & 0 & K \cr
0 & 0 & 0 \cr
K & 0 & 0
\end{pmatrix}, \cr
&K^z =
\begin{pmatrix}
0 & K & 0 \cr
K & 0 & 0 \cr
0 & 0 & 0
\end{pmatrix}.
\end{align}

%%%%%%%%%%%%%%%%%%%%%%%%%%%%%%%%%%%%%%%%%%%%%%%%%%%%%%%%%%%%%%%%%%%%%%%%%%%%%%%%%%%%%%
\begin{figure}[t]
\begin{center}
\includegraphics[width=6cm]{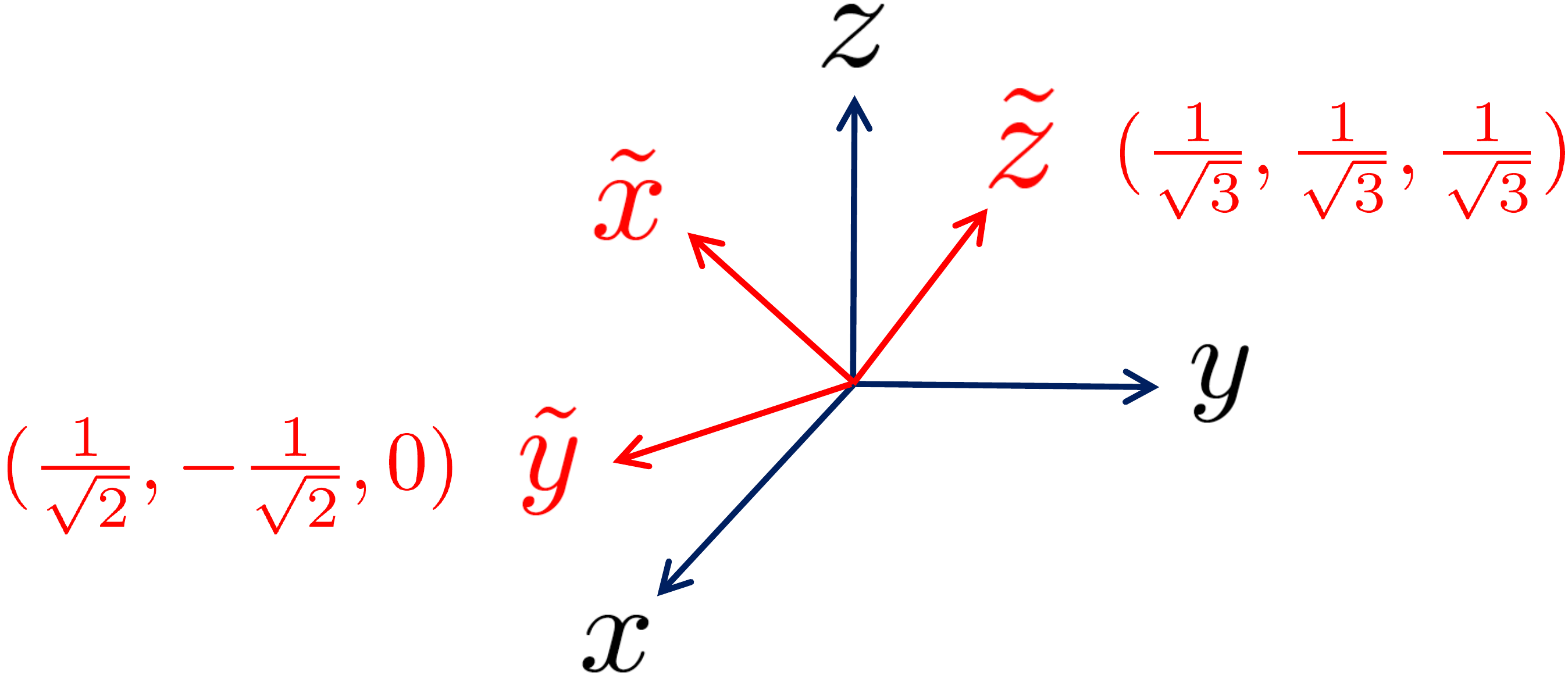}
\end{center}
\caption{
(Color online)
$\tx\ty\tz$ coordinate.
The unit vectors are taken as
$\tbe_z=(\frac{1}{\sqrt{3}},\frac{1}{\sqrt{3}},\frac{1}{\sqrt{3}})$,
$\tbe_y=(\frac{1}{\sqrt{2}},-\frac{1}{\sqrt{2}},0)$,
$\tbe_x=\tbe_y\times\tbe_z= (-\frac{1}{\sqrt{6}},-\frac{1}{\sqrt{6}},\frac{2}{\sqrt{6}})$.
}
\label{fig:cubic}
\end{figure}
%%%%%%%%%%%%%%%%%%%%%%%%%%%%%%%%%%%%%%%%%%%%%%%%%%%%%%%%%%%%%%%%%%%%%%%%%%%%%%%%%%%%%%

In the cubic symmetry, there are threefold axes along the $[111]$ and its equivalent directions.
We take the $\tz$-axis along the $[111]$ direction.
The $\tx$- and $\ty$-axes are taken perpendicular to the $\tz$-axis, as shown in Fig. \ref{fig:cubic}
We introduce the following unitary transformation $U$ whose matrix elements are given by $U_{ij} = \tbe_i\cdot\be_j$:
\begin{align}
U =
\begin{pmatrix}
-\frac{1}{ \sqrt{6}} & - \frac{1}{\sqrt{6}} & \frac{2}{\sqrt{6}} \\
\frac{1}{\sqrt{2}} & - \frac{1}{\sqrt{2}} & 0 \\
\frac{1}{\sqrt{3}} & \frac{1}{\sqrt{3}} & \frac{1}{\sqrt{3}}
\end{pmatrix}
\end{align}
We express $S^\alpha$ and $\tS^\alpha$ as the $\alpha$ component of the spin operators
in the $xyz$ and $\tx\ty\tz$ coordinates, respectively.
$S^\alpha$ is represented as
$S^\alpha=U^\dagger_{\alpha\beta} \tS^\beta$.
In the same way, $p^\alpha$ ($\tp^\alpha$) is the $\alpha$ component of the electric dipole in the $xyz$ ($\tx\ty\tz$) coordinate.
$\tp^\alpha$ is expressed as
\begin{align}
\tp^\alpha 
&= U_{\alpha\alpha'} p^{\alpha'} \cr
&= U_{\alpha\alpha'} S^{\beta'} K^{\alpha'}_{\beta'\gamma'} S^{\gamma'} \cr
&= \tS^{\beta} \left( U_{\alpha\alpha'} U_{\beta\beta'} K^{\alpha'}_{\beta'\gamma'} U^\dagger_{\gamma'\gamma} \right) \tS^{\gamma} \cr
&= \tS^{\beta} {\tK}^\alpha_{\beta\gamma} \tS^{\gamma}.
\label{eqn:p-tilde}
\end{align}
Here,
\begin{align}
\tK^\alpha_{\beta\gamma} = U_{\alpha\alpha'} U_{\beta\beta'} K^{\alpha'}_{\beta'\gamma'} U^\dagger_{\gamma'\gamma}
\end{align}
is the tensor for the electric dipole in the $\tx\ty\tz$ coordinate.
The $\alpha(=x,y,z)$ components are given by
\begin{align}
&\tK^x = K
\begin{pmatrix}
\sqrt{\frac{2}{ 3}} & 0 & -\frac{1}{\sqrt{3}} \cr
0  & - \sqrt{\frac{2}{3}} & 0 \cr
- \frac{1}{\sqrt{3}} & 0 & 0
\end{pmatrix}, \cr
&\tK^y = K
\begin{pmatrix}
0 & - \sqrt{\frac{2}{ 3}} & 0 \cr
- \sqrt{\frac{2}{3}} & 0 & - \frac{1}{\sqrt{3}} \cr
0 & - \frac{1}{\sqrt{3}} & 0
\end{pmatrix}, \cr
&\tK^z = K
\begin{pmatrix}
- \frac{1}{\sqrt{3}} & 0 & 0 \cr
0 & - \frac{1}{\sqrt{3}} & 0 \cr
0 & 0 & \frac{2}{\sqrt{3}}
\end{pmatrix}.
\label{eqn:K-tilde}
\end{align}
Substituting Eq. (\ref{eqn:K-tilde}) into Eq. (\ref{eqn:p-tilde}), we obtain
\begin{align}
\begin{pmatrix}
\tp^x \cr
\tp^y
\end{pmatrix}
&=
\begin{pmatrix}
\tO^{zx} & \tO^{x^2-y^2} \cr
\tO^{yz} & - \tO^{xy}
\end{pmatrix}
\begin{pmatrix}
- \frac{1}{\sqrt{3}} K \cr
\sqrt{\frac{2}{3}} K
\end{pmatrix}, \cr
\tp^z &= \tO^{z^2} K.
\label{eqn:p-cubic}
\end{align}
Here, $\tO^m$ ($m=yz,zx,xy,x^2-y^2,z^2$) represent the quadrupole operators in the $\tx\ty\tz$ coordinate.

%%%%%%%%%%%%%%%%%%%%%%%%%%%%%%%%%%%%%%%%%%%%%%%%%%%%%%%%%%%%%%%%%%%%%%%%%%%%%%%%%%%%%%%%%%%%%%%%%%%
\section{Directional dichroism}
\label{appendix:directional}
%%%%%%%%%%%%%%%%%%%%%%%%%%%%%%%%%%%%%%%%%%%%%%%%%%%%%%%%%%%%%%%%%%%%%%%%%%%%%%%%%%%%%%%%%%%%%%%%%%%

Let us give a short summary on appearance of the directional dichroism.
We consider the following perturbation Hamiltonian for light absorption:
\begin{align}
\H' = - g\mu_{\rm B} H^\omega S^\alpha - E^\omega P^\beta.
\end{align}
Here, $H^\omega$ and $E^\omega$ are alternating magnetic and electric fields of light, respectively.
$S^\alpha$ and $P^\beta$ are the $\alpha$ and $\beta$ components of the total spin and total electric dipole operators, respectively.
When a transition from an initial state $|i\rangle$ to a finial state $|f\rangle$ is cause by a light,
the absorption rate is proportional to
\begin{align}
|\braket{f| \left( g\mu_{\rm B} H^\omega S^\alpha + E^\omega P^\beta \right) |i}|^2.
\end{align}

%%%%%%%%%%%%%%%%%%%%%%%%%%%%%%%%%%%%%%%%%%%%%%%%%%%%%%%%%%%%%%%%%%%%%%%%%%%%%%%%%%%%%%%%%%%%%%%%%%%
\subsection{Inversion symmetry}
\label{sec:inversion}
%%%%%%%%%%%%%%%%%%%%%%%%%%%%%%%%%%%%%%%%%%%%%%%%%%%%%%%%%%%%%%%%%%%%%%%%%%%%%%%%%%%%%%%%%%%%%%%%%%%

First, we discuss the transformation of the total electric dipole with respect to the inversion operation in \Co.
The $\beta$ component of the total electric dipole operator can be written as
$P^\beta=p_a^\beta+p_{b}^\beta+p_{a'}^\beta+p_{b'}^\beta$,
where the electric dipole at each site is given by
Eqs. (\ref{eqn:p-A}), (\ref{eqn:pB-inverse}), (\ref{eqn:pB-2-fold}), and (\ref{eqn:pA'}).
The dipole is described by quadrupole operators.
They are invariant under the inversion transformation $I$ as $IO^mI^{-1}=O^m$ ($m=yz,zx,xy,x^2-y^2,z^2$),
whereas the site indexes for the quadrupole operators are interchanged as $a\leftrightarrow b$ and $a'\leftrightarrow b'$.
Since the coefficient $K$ between $p$ and $O$ is staggered between the $a\leftrightarrow b$ and $a'\leftrightarrow b'$ sites
(see Table \ref{table:matrix-T-mode}),
we obtain
$I(p_a^\beta,p_b^\beta,p_{a'}^\beta,p_{b'}^\beta)I^{-1}=(-p_b^\beta,-p_a^\beta,-p_{b'}^\beta,-p_{a'}^\beta)I^{-1}$.
This leads to the general result of $IP^\beta I^{-1}=-P^\beta$.

When the Hamiltonian is invariant under the inversion transformation, the energy eigenstates are classified by the parity as
$I|i\rangle = \lambda_i^I|i\rangle$ and $I|f\rangle = \lambda_f^I|f\rangle$.
Here, $\lambda_i^I$ and $\lambda_f^I$ are eigenvalues and take $+1$ or $-1$.
In this case, the matrix element is calculated as
\begin{align}
\braket{f| ( \H_s + \H_p ) |i}
&=\braket{If| I ( \H_s + \H_p ) I^{-1} |I i} \cr
&=\lambda_f^I \lambda_i^I \braket{f| ( \H_s - \H_p ) |i}.
\end{align}
Here, $\H_s=g\mu_{\rm B} H^\omega S^\alpha$ and $\H_p=E^\omega P^\beta$, and we used $I S^\alpha I^{-1}=S^\alpha$ and $I P^\beta I^{-1}=-P^\beta$.
Then, we obtain
\begin{align}
|\braket{f| ( \H_s + \H_p ) |i}|^2 =|\braket{f| ( \H_s - \H_p ) |i}|^2.
\label{eqn:matrix-I}
\end{align}
This means that the value is the same under
$(\H_p,\H_s)\rightarrow (-\H_p,\H_s)$ or $(\H_p,\H_s)\rightarrow (\H_p,-\H_s)$ transformations,
i.e. the absorption rate is the same when the direction of light is reversed as
$(E^\omega,H^\omega)\rightarrow (-E^\omega,H^\omega)$ or $(E^\omega,H^\omega)\rightarrow (E^\omega,-H^\omega)$.
This indicates that the directional dichroism does not appear in the presence of the inversion symmetry.
In the absence of the inversion symmetry, Eq. (\ref{eqn:matrix-I}) does not hold
and the absorption rate can be different when the direction of light is reversed.

%%%%%%%%%%%%%%%%%%%%%%%%%%%%%%%%%%%%%%%%%%%%%%%%%%%%%%%%%%%%%%%%%%%%%%%%%%%%%%%%%%%%%%%%%%%%%%%%%%%
\subsection{Time-reversal symmetry}
%%%%%%%%%%%%%%%%%%%%%%%%%%%%%%%%%%%%%%%%%%%%%%%%%%%%%%%%%%%%%%%%%%%%%%%%%%%%%%%%%%%%%%%%%%%%%%%%%%%

When the Hamiltonian is invariant under the time-reversal transformation $\Theta$,
the energy eigenstates are also eigenstates of $\Theta$ as
$\Theta |i\rangle = \lambda_i^\Theta |i\rangle$ and $\Theta |f\rangle = \lambda_f^\Theta |f\rangle$.
Here, $\lambda_i^\Theta$ and $\lambda_f^\Theta$ are eigenvalues for $|i\rangle$ and $|f\rangle$, respectively.
They are complex and satisfy $|\lambda_i^\Theta|=|\lambda_f^\Theta|=1$.
Here, we assumed that the $|i\rangle$ and $|f\rangle$ states have no degeneracy.
The matrix element is calculated as
\cite{J-J-Sakurai-2017}
\begin{align}
&\braket{f| ( \H_s + \H_p ) |i}
=\braket{\tilde{i}| \Theta ( \H_s + \H_p ) \Theta^{-1} |\tilde{f}} \cr
&~~~~~~~~~~~~~~~~~~
=\left(\lambda_i^\Theta\right)^* \lambda_f^\Theta \braket{i| ( - \H_s + \H_p ) |f}.
\label{eqn:directional-T}
\end{align}
Here, we introduced $|\tilde{i}\rangle=\Theta |i\rangle$ and $|\tilde{f}\rangle=\Theta |f\rangle$,
and used $\Theta S^\alpha \Theta^{-1}=-S^\alpha$ and $\Theta P^\beta \Theta^{-1}=P^\beta$.
Then, we obtain
\begin{align}
|\braket{f| ( \H_s + \H_p ) |i}|^2 =|\braket{f| ( - \H_s + \H_p ) |i}|^2.
\label{eqn:matrix-T}
\end{align}
The value is the same under the $(\H_p,\H_s)\rightarrow(-\H_p,\H_s)$ or $(\H_p,\H_s)\rightarrow(\H_p,-\H_s)$ transformations
and the directional dichroism does not appear in the presence of the time-reversal symmetry.

%%%%%%%%%%%%%%%%%%%%%%%%%%%%%%%%%%%%%%%%%%%%%%%%%%%%%%%%%%%%%%%%%%%%%%%%%%%%%%%%%%%%%%
\begin{table}[t]
\caption{
Transformations of electromagnetic fields by Eqs. (\ref{eqn:trans-1}) and (\ref{eqn:trans-2}) for $C_{2y}$ and $\sigma_y^c\Theta$, respectively.
We consider circularly polarized light here.
We take amplitudes of the fields as unity.
$(\pm x,{\rm L})$ represent the left circularly polarized lights propagating along the $\pm x$ directions, respectively.
$(\pm x,{\rm R})$ are for the right circularly polarized light.
The lights propagating in the $y$ and $z$ directions are expressed in the same way.
Let us consider a light propagating in the $x$ direction, for instance.
For the initial field, we consider the left circularly polarized light here.
Notice that the same argument also holds for the right circularly polarized light by interchanging ${\rm L}\leftrightarrow{\rm R}$.
In the presence of $C_{2y}$, $(+x,{\rm L})$ is transformed into $(-x,{\rm L})$.
This means that the absorption rate is invariant as $W_{\rm L}(+x)=W_{\rm L}(-x)$ and magnetic circular dichroism (MCD) does not appear.
Definitions of various types of dichroism are given in the caption of Table \ref{table:dichroism}.
In the presence of $\sigma_y^c\Theta$, $(+x,{\rm L})$ is transformed into $(-x,{\rm R})$.
This means that the absorption rate is invariant as $W_{\rm L}(+x)=W_{\rm R}(-x)$ and natural circular dichroism (NCD) does not appear.
In the presence of both $C_{2y}$ and $\sigma_y^c\Theta$, $(+x,{\rm L})$ can be transformed into $(+x,{\rm R})$.
This assures $W_{\rm L}(+x)=W_{\rm R}(+x)$ and circular dichroism (CD) does not appear.
For a light propagating in the $y$ direction,
there is no restriction in the absorption rate with respect to the $\pm$ directions of the propagation and both MCD and NCD can appear.
In the presence of $\sigma_y^c\Theta$, on the other hand, $W_{\rm L}(+y)=W_{\rm R}(+y)$ and CD does not appear.
}
\begin{tabular}{ccccc}
\hline
Electromagnetic Fields & Initial & $C_{2y}$ & $\sigma_y^c\Theta$ \cr
\hline
$x$-propagation & $(+x,{\rm  L})$ & $(-x,{\rm L})$ & $(-x,{\rm R})$ \cr
$\bE^\omega$ & $(0,1,i)$ & $(0,1,-i)$ & $(0,1,i)$ \cr
$\bH^\omega$ & $(0,-i,1)$ & $(0,-i,-1)$ & $(0,i,-1)$ \cr
\hline
$y$-propagation & $(+y,{\rm L})$ & $(+y,{\rm L})$ & $(+y,{\rm R})$ \cr
$\bE^\omega$ & $(i,0,1)$ & $(-i,0,-1)$ & $(i,0,-1)$ \cr
$\bH^\omega$ & $(1,0,-i)$ & $(-1,0,i)$ & $(-1,0,-i)$ \cr
\hline
$z$-propagation & $(+z,{\rm L})$ & $(-z,{\rm L})$ & $(-z,{\rm R})$ \cr
$\bE^\omega$ & $(1,i,0)$ & $(-1,i,0)$ & $(-1,-i,0)$ \cr
$\bH^\omega$ & $(-i,1,0)$ & $(i,1,0)$ & $(-i,1,0)$ \cr
\hline
\end{tabular}
\label{table:trans-circular}
\end{table}
%%%%%%%%%%%%%%%%%%%%%%%%%%%%%%%%%%%%%%%%%%%%%%%%%%%%%%%%%%%%%%%%%%%%%%%%%%%%%%%%%%%%%%

When the $|i\rangle$ and $|f\rangle$ states have degeneracy, as in half-integer spin cases,
the $|\tilde{i}\rangle$ and $|\tilde{f}\rangle$ states are their Kramers' partners, respectively.
In this case, we consider the following matrix elements for the degenerate initial and final states:
\begin{align}
&|\braket{f| ( \H_s + \H_p ) |i}|^2 + |\braket{\tilde{f}| ( \H_s + \H_p ) |\tilde{i}}|^2 \cr
&~~~
+ |\braket{\tilde{f}| ( \H_s + \H_p ) |i}|^2 + |\braket{f| ( \H_s + \H_p ) |\tilde{i}}|^2 \cr
&=
|\braket{f| ( \H_s + \H_p ) |i}|^2 + |\braket{i| \Theta ( \H_s + \H_p ) \Theta^{-1}|f}|^2 \cr
&~~~
+ |\braket{\tilde{f}| ( \H_s + \H_p ) |i}|^2 + |\braket{i| \Theta ( \H_s + \H_p ) \Theta^{-1} |\tilde{f}}|^2 \cr
&=
|\braket{f| ( \H_s + \H_p ) |i}|^2 + |\braket{f| ( - \H_s + \H_p ) |i}|^2 \cr
&~~~
+ |\braket{\tilde{f}| ( \H_s + \H_p ) |i}|^2 + |\braket{\tilde{f}| ( - \H_s + \H_p ) |i}|^2.
\label{eqn:matrix-T-2}
\end{align}
The value is the same under the $(\H_p,\H_s)\rightarrow (-\H_p,\H_s)$ or $(\H_p,\H_s)\rightarrow (\H_p,-\H_s)$ transformations
and the directional dichroism does not appears, as in the nondegenerate case.

Thus, breaking of both the inversion and time-reversal symmetries are required for the appearance of the directional dichroism.

%%%%%%%%%%%%%%%%%%%%%%%%%%%%%%%%%%%%%%%%%%%%%%%%%%%%%%%%%%%%%%%%%%%%%%%%%%%%%%%%%%%%%%%%%%%%%%%%%%%
\section{Transformations of Electromagnetic Fields Under $C_{2y}$ and $\sigma_y^c\Theta$ Symmetries}
\label{appendix:trans}
%%%%%%%%%%%%%%%%%%%%%%%%%%%%%%%%%%%%%%%%%%%%%%%%%%%%%%%%%%%%%%%%%%%%%%%%%%%%%%%%%%%%%%%%%%%%%%%%%%%

In the presence of $C_{2y}$, the absorption rate of light is invariant under the transformation of electromagnetic fields given by Eq. (\ref{eqn:trans-1}).
Similarly, in the presence of $\sigma_y^c\Theta$, the absorption rate is invariant under the transformation given by Eq. (\ref{eqn:trans-2}).
In Tables \ref{table:trans-circular} and \ref{table:trans-linear}, we list the transformed electromagnetic fields for circularly and linearly polarized lights, respectively.

%%%%%%%%%%%%%%%%%%%%%%%%%%%%%%%%%%%%%%%%%%%%%%%%%%%%%%%%%%%%%%%%%%%%%%%%%%%%%%%%%%%%%%
\begin{table}[t]
\caption{
Transformations of electromagnetic fields by Eqs. (\ref{eqn:trans-1}) and (\ref{eqn:trans-2}) for $C_{2y}$ and $\sigma_y^c\Theta$, respectively.
We consider linearly polarized light here.
In this case, the results becomes the same between $C_{2y}$ and $\sigma_y^c\Theta$.
We take amplitudes of the fields as unity.
$\pm x$, $\pm y$, and $\pm z$ represent lights propagating along the corresponding directions.
$\theta$ represents the direction of a linearly polarized light.
When a light propagates in the $x$ direction, the $+x$ propagation is transformed into the $-x$ one.
However, the electromagnetic fields are not equivalent after the transformation.
This means that the absorption rates can be different between the $+x$ and $-x$ propagations.
On the other hand, the electromagnetic fields become equivalent for specific values of $\theta$ as $\theta=0,\pm\frac{\pi}{2},\pi$.
For these directions of the polarization, the absorption rate is invariant as $W(+x)=W(-x)$ and directional dichroism (DD) does not appear.
When the light propagates in the $y$ direction, there is no restriction in the absorption rate and DD can appear in any directions of the polarization.
}
\begin{tabular}{cccc}
\hline
Electromagnetic Fields & Initial & $C_{2y},~\sigma_y^c\Theta$ \cr
\hline
$x$-propagation & $+x$ & $-x$ \cr
$\bE^\omega$ & $(0,\cos\theta,-\sin\theta)$ & $(0,\cos\theta,\sin\theta)$ \cr
$\bH^\omega$ & $(0,\sin\theta,\cos\theta)$ & $(0,\sin\theta,-\cos\theta)$ \cr
\hline
$y$-propagation & $+y$ & $+y$ \cr
$\bE^\omega$ & $(-\sin\theta,0,\cos\theta)$ & $(\sin\theta,0,-\cos\theta)$ \cr
$\bH^\omega$ & $(\cos\theta,0,\sin\theta)$ & $(-\cos\theta,0,-\sin\theta)$ \cr
\hline
$z$-propagation & $+z$ & $-z$ \cr
$\bE^\omega$ & $(\cos\theta,-\sin\theta,0)$ & $(-\cos\theta,-\sin\theta,0)$ \cr
$\bH^\omega$ & $(\sin\theta,\cos\theta,0)$ & $(-\sin\theta,\cos\theta,0)$ \cr
\hline
\end{tabular}
\label{table:trans-linear}
\end{table}
%%%%%%%%%%%%%%%%%%%%%%%%%%%%%%%%%%%%%%%%%%%%%%%%%%%%%%%%%%%%%%%%%%%%%%%%%%%%%%%%%%%%%%

In case of a left circularly polarized light propagating in the $+z$ direction, for instance, we express the electromagnetic fields as
\begin{align}
&\bE = 2E^\omega \left(\cos(kz-\omega t),-\sin(kz-\omega t),0\right), \cr
&\bH = 2H^\omega \left(\sin(kz-\omega t),\cos(kz-\omega t),0\right).
\end{align}
For the later convenience, we took $2E^\omega$ and $2H^\omega$ as the amplitudes of the electric and magnetic fields, respectively.
When we consider a light absorption process, the cosine and sine terms are treated as
$\cos(kz-\omega t)\rightarrow \frac{1}{2}e^{-i\omega t}$ and
$\sin(kz-\omega t)\rightarrow -i\frac{1}{2}e^{-i\omega t}$.
Here, we assumed that the wavelength is much larger than the sample size and neglected the spatial dependent $kz$ term.
For the circularly polarized light, the electromagnetic fields in the matrix element in Eq. (\ref{eqn:matrix-00}) are expressed as
\begin{align}
\bE^\omega = E^\omega (1,i,0),~~~
\bH^\omega = H^\omega (-i,1,0).
\end{align}
This expression is used in Table \ref{table:trans-circular}.

%%%%%%%%%%%%%%%%%%%%%%%%%%%%%%%%%%%%%%%%%%%%%%%%%%%%%%%%%%%%%%%%%%%%%%%%%%%%%%%%%%%%%%%%%%%%%%%%%%%
\section{Matrices of $S=3/2$ Spin Operators}
\label{appendix:spin}
%%%%%%%%%%%%%%%%%%%%%%%%%%%%%%%%%%%%%%%%%%%%%%%%%%%%%%%%%%%%%%%%%%%%%%%%%%%%%%%%%%%%%%%%%%%%%%%%%%%

The matrix forms of the $S=3/2$ spin operators are expressed as
\begin{align}
&S^x =
\begin{pmatrix}
0 & \frac{\sqrt{3}}{2} & 0 & 0 \cr
\frac{\sqrt{3}}{2} & 0 & 1 & 0 \cr
0 & 1 & 0 & \frac{\sqrt{3}}{2} \cr
0 & 0 & \frac{\sqrt{3}}{2} & 0 \cr
\end{pmatrix}, \cr
&S^y =
\begin{pmatrix}
0 & -i\frac{\sqrt{3}}{2} & 0 & 0 \cr
i \frac{\sqrt{3}}{2} & 0 & -i & 0 \cr
0 & i & 0 & -i\frac{\sqrt{3}}{2} \cr
0 & 0 & i\frac{\sqrt{3}}{2} & 0 \cr
\end{pmatrix}, \cr
&S^z =
\begin{pmatrix}
\frac{3}{2} & 0 & 0 & 0 \cr
0 & \frac{1}{2} & 0 & 0 \cr
0 & 0 & -\frac{1}{2} & 0 \cr
0 & 0 & 0 & -\frac{3}{2}
\end{pmatrix}.
\label{eqn:S-mat}
\end{align}

%%%%%%%%%%%%%%%%%%%%%%%%%%%%%%%%%%%%%%%%%%%%%%%%%%%%%%%%%%%%%%%%%%%%%%%%%%%%%%%%%%%%%%%%%%%%%%%%%%%
\section{Matrices of Quadrupole Operators for $S=3/2$}
\label{appendix:quadrupole}
%%%%%%%%%%%%%%%%%%%%%%%%%%%%%%%%%%%%%%%%%%%%%%%%%%%%%%%%%%%%%%%%%%%%%%%%%%%%%%%%%%%%%%%%%%%%%%%%%%%

The quadrupole operators are expressed as
\begin{align}
&O^{yz} = S^y S^z + S^z S^y
= \sqrt{3}
\begin{pmatrix}
0 & -i & 0 & 0 \cr
i & 0 & 0 & 0 \cr
0 & 0 & 0 & i \cr
0 & 0 & -i & 0
\end{pmatrix}, \cr
&O^{zx} = S^z S^x + S^x S^z
= \sqrt{3}
\begin{pmatrix}
0 & 1 & 0 & 0 \cr
1 & 0 & 0 & 0 \cr
0 & 0 & 0 & -1 \cr
0 & 0 & -1 & 0
\end{pmatrix}, \cr
&O^{xy} = S^x S^y + S^y S^x
= \sqrt{3}
\begin{pmatrix}
0 & 0 & -i & 0 \cr
0 & 0 & 0 & -i \cr
i & 0 & 0 & 0 \cr
0 & i & 0 & 0
\end{pmatrix}, \cr
&O^{x^2-y^2} = (S^x)^2 - (S^y)^2
= \sqrt{3}
\begin{pmatrix}
0 & 0 & 1 & 0 \cr
0 & 0 & 0 & 1 \cr
1 & 0 & 0 & 0 \cr
0 & 1 & 0 & 0
\end{pmatrix}, \cr
&O^{z^2} = \frac{1}{\sqrt{3}} \left[3(S^z)^2 - \bS^2 \right]
= \sqrt{3}
\begin{pmatrix}
1 & 0 & 0 & 0 \cr
0 & -1 & 0 & 0 \cr
0 & 0 & -1 & 0 \cr
0 & 0 & 0 & 1
\end{pmatrix}. \cr
\label{eqn:O-mat}
\end{align}

%%%%%%%%%%%%%%%%%%%%%%%%%%%%%%%%%%%%%%%%%%%%%%%%%%%%%%%%%%%%%%%%%%%%%%%%%%%%%%%%%%%%%%%%%%%%%%%%%%%

%%%%%%%%%%%%%%%%%%%%%%%%%%%%%%%%%%%%%%%%%%%%%%%%%%%%%%%%%%%%%%%%%%%%%%%%%%%%%%%%%%%%%%%%%%%%%%%%%%%


\begin{thebibliography}{99}
\bibitem{note-c-glide}
The $c$-glide symmetry can be checked by subsequent operations of the inversion and twofold transformations shown in Fig. \ref{fig:crystal}.
Let us begin with a position at $(x,y,z)_1$.
After the inversion operation at $(0,0,0)$, we obtain $(x,y,z)_1\rightarrow(-x,-y,-z)_2$.
With respect to the twofold operation at $(0,0,\frac{1}{4})$, we obtain $(-x,-y,-z)_2\rightarrow(x,-y,z+\frac{1}{2})_3$.
It is transformed by the inversion at $(0,0,\frac{1}{2})$ as $(x,-y,z+\frac{1}{2})_3\rightarrow(-x,y,-z+\frac{1}{2})_4$.
As for the transformation of $(x,y,z)_1\rightarrow(x,-y,z+\frac{1}{2})_3$,
it is realized by a $c$-glide with respect to the mirror operation perpendicular to the $y$ direction with a shift of ($0,0,\frac{1}{2})$.
Similarly, the transformation $(-x,-y,-z)_2\rightarrow(-x,y,-z+\frac{1}{2})_4$ is also realized by the $c$-glide.
\bibitem{note-P-C2}
This can be checked in parallel with the discussion in Appendix \ref{sec:inversion}.
With respect to the $C_{2y}$ operation, the quadrupole operators are transformed as
$C_{2y}(O^{yz},O^{zx},O^{xy},O^{x^2-y^2},O^{z^2})C_{2y}^{-1}=(-O^{yz},O^{zx},-O^{xy},O^{x^2-y^2},O^{z^2})$,
whereas the site indexes for the quadrupole operators are interchanged as $a\leftrightarrow b'$ and $b\leftrightarrow a'$.
This leads to $C_{2y}(P^x,P^y,P^z)C_{2y}^{-1}=(-P^x,P^y,-P^z)$.
\bibitem{note-P-c-glide}
This can be checked in parallel with the discussion in Appendix \ref{sec:inversion}.
With respect to the mirror operation perpendicular to the $y$ direction ($\sigma_y$),
the quadrupole operators are transformed as
$\sigma_y(O^{yz},O^{zx},O^{xy},O^{x^2-y^2},O^{z^2})\sigma_y^{-1}=(-O^{yz},O^{zx},-O^{xy},O^{x^2-y^2},O^{z^2})$.
The subsequent translation along the $z$ direction with $(0,0,\frac{1}{2})$
interchanges the site indexes for the quadrupole operators as $a\leftrightarrow a'$ and $b\leftrightarrow b'$.
With respect to the $c$-glide operation ($\sigma_y^c$),
we obtain $\sigma_y^c(P^x,P^y,P^z)(\sigma_y^c)^{-1}=(P^x,-P^y,P^z)$.
\bibitem{note-classification}
Let us assume that operators $A_\pm$ satisfy $C_{2x} A_\pm C_{2x}^{-1}=\pm A_\pm$, respectively.
$|\alpha\rangle$ and $|\beta\rangle$ are eigenstates of $C_{2x}$ as
$C_{2x}|\alpha\rangle=\lambda_\alpha|\alpha\rangle$ and $C_{2x}|\beta\rangle=\lambda_\beta|\beta\rangle$.
The following matrix element is calculated as
$\langle\beta|A_\pm|\alpha\rangle=\langle C_{2x} \beta|C_{2x}A_\pm C_{2x}^{-1}|C_{2x}\alpha\rangle
=\pm\lambda_\beta^* \lambda_\alpha \langle\beta|A_\pm|\alpha\rangle.$
When $\langle\beta|A_+|\alpha\rangle\neq 0$, $\lambda_\beta^* \lambda_\alpha=1$
and this leads to $\langle\alpha|A_-|\beta\rangle=0$.
When $\langle\beta|A_-|\alpha\rangle\neq 0$, $\lambda_\beta^* \lambda_\alpha=-1$
and this leads to $\langle\alpha|A_+|\beta\rangle=0$.
Therefore, the $A_\pm$ operators cannot connect the common states
and they are classified in different groups, as shown in Fig. \ref{fig:level-hx}.
\bibitem{note-sign-hx}
\label{ref:note-sign-hx}
Let us begin with the following eigenstate defined by $\H_0(h_{\rm MF})|m\rangle=E_m|m\rangle$
with $\H_0(h_{\rm MF})=D(S^z)^2-h_{\rm MF}S^x-h_z S^z$ as the local Hamiltonian.
Here, $h_{\rm FM}$ and $h_z$ represent the molecular and external magnetic fields, respectively.
The molecular field $h_{\rm MF}$ in $\H_0(h_{\rm MF})$ can be reversed by the following transformation:
$\H_0(-h_{\rm MF})=C_{2z}H_0(h_{\rm MF})C_{2z}^{-1}$.
Here, $C_{2z}$ represents the twofold rotational operation around the $z$-axis.
Using this transformation, we can obtain
\begin{align*}
C_{2z}\H_0(h_{\rm MF}) C_{2z}^{-1} C_{2z} |m\rangle &= E_m C_{2z}|m\rangle \cr
\H_0(-h_{\rm MF}) |\tilde{m}\rangle &= E_m |\tilde{m}\rangle.
\end{align*}
Here, $|\tilde{m}\rangle=C_{2z}|m\rangle$ is the energy eigenstate with $E_m$ for $\H_0(-h_{\rm MF})$.
The spin and quadrupole operators are transformed as
$C_{2z}(S^x,S^y,S^z)C_{2z}^{-1}=(-S^x,-S^y,S^z)$
and
$C_{2z}(O^{yz},O^{zx},O^{xy},O^{x^2-y^2},O^{z^2})C_{2z}^{-1}=(-O^{yz},-O^{zx},O^{xy},O^{x^2-y^2},O^{z^2})$.
The matrix elements are calculated as
\begin{align*}
&\braket{l|(S^x,S^y,S^z)|m} \cr
&~~~
= \braket{C_{2z}l| C_{2z}(S^x,S^y,S^z)C_{2z}^{-1}|C_{2z}m} \cr
&~~~
= \braket{\tilde{l}| (-S^x,-S^y,S^z)|\tilde{m}}, \cr
&\braket{l|(O^{yz},O^{zx},O^{xy},O^{x^2-y^2},O^{z^2})|m} \cr
&~~~
= \braket{C_{2z}l| C_{2z}(O^{yz},O^{zx},O^{xy},O^{x^2-y^2},O^{z^2})C_{2z}^{-1}|C_{2z}m} \cr
&~~~
= \braket{\tilde{l}|(-O^{yz},-O^{zx},O^{xy},O^{x^2-y^2},O^{z^2})|\tilde{m}}.
\end{align*}
Here, $|\tilde{l}\rangle=C_{2z}|l\rangle$ and $|\tilde{m}\rangle=C_{2z}|m\rangle$ were introduced.
These results represent the relation of the matrix element between $\H_0(h_{\rm MF})$ (A site) and $\H_0(-h_{\rm MF})$ (B site).
They hold in both $\bm{H}=0$ and $\bm{H}\parallel z$ cases.
\end{thebibliography}
\end{document}